\documentclass[aps,prb,twocolumn,superscriptaddress,floatfix,longbibliography,citeautoscript]{revtex4-2}

\usepackage{amsmath,amssymb} 
\usepackage{bm} 
\usepackage{graphicx} 
\usepackage{comment} 

\usepackage{tikz}

\usepackage{braket}
\usepackage{circuitikz}

\usepackage{textcomp} 

\usepackage{enumitem}
\setlist{noitemsep,leftmargin=*,topsep=0pt,parsep=0pt}

\usepackage{xcolor} 
\definecolor{lightgray}{gray}{0.6}
\definecolor{medgray}{gray}{0.4}

\usepackage{hyperref}
\hypersetup{
colorlinks=true,
urlcolor= blue,
citecolor=blue,
linkcolor= blue,
}

\definecolor{darkblue}{rgb}{0.0, 0.0, 0.75}

\usepackage[normalem]{ulem} 
\definecolor{mgreen}{RGB}{1,123,0}

\def \br{{\bf r}}

\def \ms{\text{s}}

\def \tV{\tilde{V} }

\def \tsigma{\tilde{\sigma} }

\def \ms{\mathrm{s}}
\def \mm{\mathrm{m}}

\newif\ifptitle
\newif\ifpnumber
\newcounter{para}

\renewcommand{\vec}[1]{\bm{#1}}

\newcommand{\uae}{Quantum Research Centre, Technology Innovation Institute, Abu Dhabi, UAE}
\newcommand{\zoq}{Zentrum für Optische Quantentechnologien and Institut für Quantenphysik, Universität Hamburg, 22761 Hamburg, Germany}
\newcommand{\cui}{The Hamburg Centre for Ultrafast Imaging, Luruper Chaussee 149, Hamburg 22761, Germany}

\newcommand{\mytitle}{Designing Atomtronic Circuits via Superfluid Dynamics}

\begin{document}

%

\title{\mytitle}

\author{Sarah Jährling}
\affiliation{\zoq} 

\author{Vijay Pal Singh}
\affiliation{\uae}

\author{Ludwig Mathey}
\affiliation{\zoq}
\affiliation{\cui}

\date{\today}

\begin{abstract}
We propose to design atomtronic circuits with Bose-Einstein condensates (BECs) in circuit-like traps that are controlled via mobile barriers. 
Using classical-field simulations, we demonstrate a universal set of logical gates and show how to assemble them into circuits. 
We first demonstrate an AND gate based on a T-shaped BEC, utilizing a combination of mobile and static barriers. 
The mobile barriers provide the logical input of the gate, while the static barrier functions as a Josephson junction 
that generates the AND output of the gate via a density imbalance across the barrier. 
Next we show how to combine three AND gates into a circuit, with a design composed of two T-shapes and an H-shape. 
Furthermore, we demonstrate how to use Josephson oscillations to create a NOT gate and combine it with an AND gate, 
thereby showcasing a universal set of gates and their assembly into circuits.
\end{abstract}

\maketitle

\section{\label{sec:Start}Introduction} 

The emergent technology of atomtronics utilizes the substantial achievements of cold-atom technology to emulate electronic circuitry \cite{Amico2021, Amico2022, Polo2024}. 
Bose-Einstein condensates (BECs), due to their low susceptibility to decoherence and dissipative effects, are particularly well-suited for studying superfluid phenomena in ultracold quantum systems \cite{Raman1999, Onofrio2000, Dalibard2000, KetterleVortex, Greiner2002,Hadzibabic2006, Salomon2014, Weimer2015, Shin2018, Pfau2018, Murthy2015, Sunami2022, Sunami2024, Rydow2024}. 
This has inspired the design of systems that closely resemble superconducting circuit elements \cite{Seaman2007,Stickney2007,Pepino2009,Pepino2010,Amico2014}.

A fundamental component in superconducting circuits is the Josephson junction (JJ), typically formed by
a superconductor-insulator-superconductor interface, 
which gives rise to the dc-ac Josephson effect \cite{Josephson1962}. 
This effect produces a supercurrent across the junction, entirely controlled by the phase difference between the two superconductors. When the current exceeds a critical current, the junction transitions into a resistive state, associated with the creation of elementary excitations \cite{barone1983josephson}. 
In cold-atom systems, atomic JJs have been realized  with static or mobile tunnel junctions, 
drawing close parallels to their superconducting counterparts.
This has enabled the measurement of current-chemical potential and current-phase relations \cite{Levy2007, Kwon2020, Luick2020, Pace2021}, and has further led to the development of atomic analogs of superconducting quantum interference devices (SQUIDs) \cite{Ramanathan2011, Wright2013, Ryu2013, Eckel2014, Ryu2020, Kiehn2022}. 
A notable extension of this analogy is the recent observation of Shapiro steps in cold-atom systems  \cite{singh2023shapiro, Del_Pace2024, Bernhart2024}.

In this paper, we propose an experimental methodology to realize a universal set of logical gates using atomic JJs , and their assembly into circuits.
We begin by constructing a two-input AND gate using a T-shaped condensate, 
which includes two mobile and one static tunnel barrier. 
The mobile barriers generate external atomic currents, 
while the accumulated density imbalance across the static barrier functions as the circuit's output.
When only one barrier is moved, no significant imbalance is observed.
However, when both barriers are moved simultaneously, a substantial imbalance is produced, 
demonstrating the successful operation of the atomic AND gate. 
To expand this approach, we design a circuit for a four-input AND gate by combining two T-shaped condensates into an H-shaped configuration. 
We demonstrate the protocol for manipulating the barriers, 
ensuring that a significant imbalance occurs only when all four input barriers are moved simultaneously. 
For all other input combinations, the resulting imbalance is small, as expected for a four-input AND gate.
Finally, we demonstrate the operation of a NOT gate using a single static barrier in a rectangular-shaped cloud, where Josephson oscillations are used to flip the input state.
With the implementation of both an AND gate and a NOT gate, we have a universal set of logical gates that can be combined to perform any classical logical operation, as demonstrated by the realization of a NAND gate.
The results are obtained using a classical-field approach, which incorporates fluctuating bosonic fields to capture dynamics beyond the mean-field approximation \cite{Blakie2008, Polkovnikov2010, Singh2016}.

This paper is structured as follows: 
In Sec. \ref{sec:SimulationMethod}, we describe the classical-field method used to simulate the dynamics of two-dimensional condensates. 
In Sec. \ref{sec:2InputAndGate}, we demonstrate the implementation of a logical two-input AND gate that exhibits a transistor-like behavior. 
In Sec. \ref{sec:4InputAndGate}, we expand this setup to construct a logical four-input AND gate. In Sec. \ref{sec:UniversalSetOfGate}, we implement a NOT gate and combine it with an AND gate to create a NAND gate, which forms the basis for a universal set of logical gates. 
We conclude in Sec. \ref{sec:Conclusions}.


%
\section{\label{sec:SimulationMethod}Simulation Method}
We consider a 2D bosonic cloud spatially confined in trap geometries that are designed to support atomtronic operation.
To simulate the dynamics we use a classical-field method within the truncated Wigner approximation \cite{Singh2016, Singh2021}. The system is described by the Hamiltonian
\begin{equation} \label{eq:hamContinuum}
\hat{H}_0 = \int d\vec{r}\bigg[ \frac{\hbar^2}{2m} \vec{\nabla} 
    \hat{\psi}^\dagger  (\vec{r}) \vec{\nabla} \hat{\psi} (\vec{r})  + 
    \frac{g}{2} \hat{\psi}^\dagger (\vec{r})\hat{\psi}^\dagger (\vec{r})\hat{\psi} (\vec{r})\hat{\psi} (\vec{r})\bigg],
\end{equation}
where $\hat{\psi}^\dagger(\vec{r})$ and $\hat{\psi} (\vec{r})$ are the creation and annihilation field operators, respectively. The interaction is given by $g = \sqrt{8\pi} a_s \hbar^2/ (l_z m)$, 
where $a_s$ is the s-wave scattering length, $l_z = \sqrt{\hbar/(m\omega_z)}$ is the harmonic oscillator length in the transverse direction, and $m$ is the atomic mass.
While we present our results for the concrete realization provided by $^6\text{Li}_2$ molecules, 
our proposed circuits can be realized with any cold-atom degenerate gas. 
We use $\tilde{g}=0.1$, the 2D density $n\approx 2.3\,\mu \text{m}^{-2}$ and the temperature $T/T_0=0.1$, 
with $g=\tilde{g} \hbar^2/m$ and  $T_0 = 2\pi n\hbar^2/(m k_B D_c)$ being the estimate of the critical temperature based on the critical phase-space density $D_c = \ln(380/\tilde{g})$ \cite{Prokof_ev_2002, PhysRevLett.87.270402}.

%
%

To perform numerical simulations, we discretize space on a lattice of size $N_x \times N_y$ with a discretization length $l = 0.5\, \mu \mathrm{m}$. The value of $l$ is chosen to be larger than both the healing length $\xi = \hbar / \sqrt{2mgn}$ and the de Broglie wavelength, ensuring that the continuum limit is satisfied \cite{PhysRevA.67.053615}. 
This procedure maps the continuum Hamiltonian onto the discrete Bose-Hubbard model, 
where the tunneling energy is defined as $J = \hbar^2/(2ml^2)$ and the on-site interaction energy as $U=g/l^2$. 
In our classical-field representation we replace the operators $\hat{\psi}$ in Eq. \ref{eq:hamContinuum} and in the equations of motion by complex numbers $\psi$. 
The initial states $\psi(\vec{r},t=0)$ are sampled in a grand canonical ensemble with chemical potential $\mu$ and temperature $T$ via a classical Metropolis algorithm.
Finally, each initial state is propagated using classical equations of motion
\begin{align}\label{eq:eom}
i \hbar \partial_t \psi_j  = - J \sum_{i(j)} \psi_{i(j)} +  V_\text{ext}(\br_j, t) \psi_j  + U |\psi_j|^2 \psi_j,
\end{align}
where $V_{\text{ext}}({\bf r},t)$ is the external barrier potential. 
The site $i(j)$ refer to the four neighboring sites $i$ of site $j$. 
To create atomtronic logical gates we use combinations of static and mobile barriers of the form 
\begin{align}
V_\ms(r,t) &= V_{0, \ms}(t) \exp{\bigg(-\frac{2 (r- r_{0, \ms})^2}{\sigma_{\ms}^2}\bigg)}, 
    \label{eq:BarrierPotentialStat}  \\  
V_\mm(r,t) &= V_{0, \mm}(t) \exp{\bigg(-\frac{2 (r- r_{0, \mm}-  v t)^2}{\sigma_{\mm}^2}\bigg)}.
    \label{eq:BarrierPotentialMob}
\end{align}
As we describe in detail below, $V_0(t)$ is the time dependent strength, $\sigma$ is the width, 
and $r_0$ is the initial position of the barrier. 
For the mobile barrier, $v$ represents its velocity. 
We define the scaled parameters $\tV_0=V_0/J$ and $\tsigma=\sigma/l$. 
We linearly ramp up $V_0(t)$ to the desired value over $150$ ms, then allow the system to relax for $50$ ms to ensure equilibrium. These ramp-up and waiting times remain fixed throughout the paper, unless stated otherwise. 
The inclusion of barriers creates weak links at $r_{0,\ms}$ and $r_{0,\mm}$, see, e.g. \cite{Singh2020}.  
To generate an atom current, we move the mobile barrier at a constant velocity $v$ 
until it reaches the final position $r_f$ \cite{Smerzi1997}.
The mobile barriers create currents, which function as input channels, 
while the density imbalance at static barriers serves as output channels. 
The behavior of these channels depends on the critical dynamics across the weak links, 
which form atomic JJs.  
Each proposed circuit element is designed based on a specific barrier protocol, as described in the following sections.

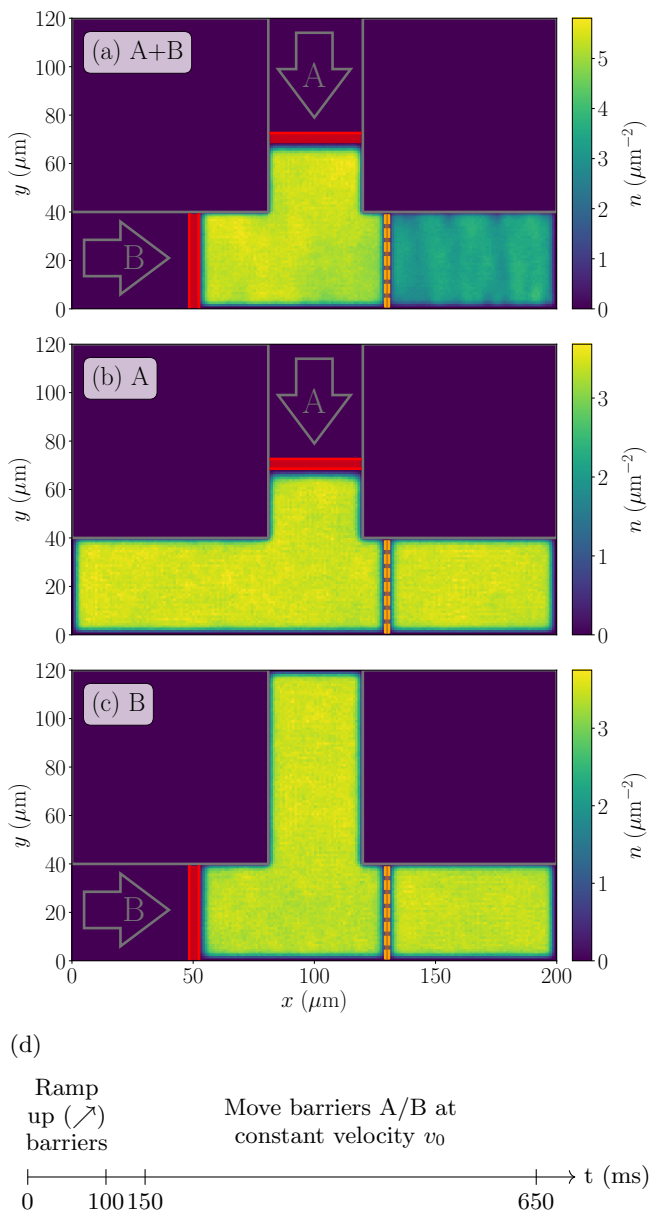
\begin{figure}[t!]
    \centering
    \begin{minipage}{\columnwidth}
        \centering
        \includegraphics[width=\textwidth]{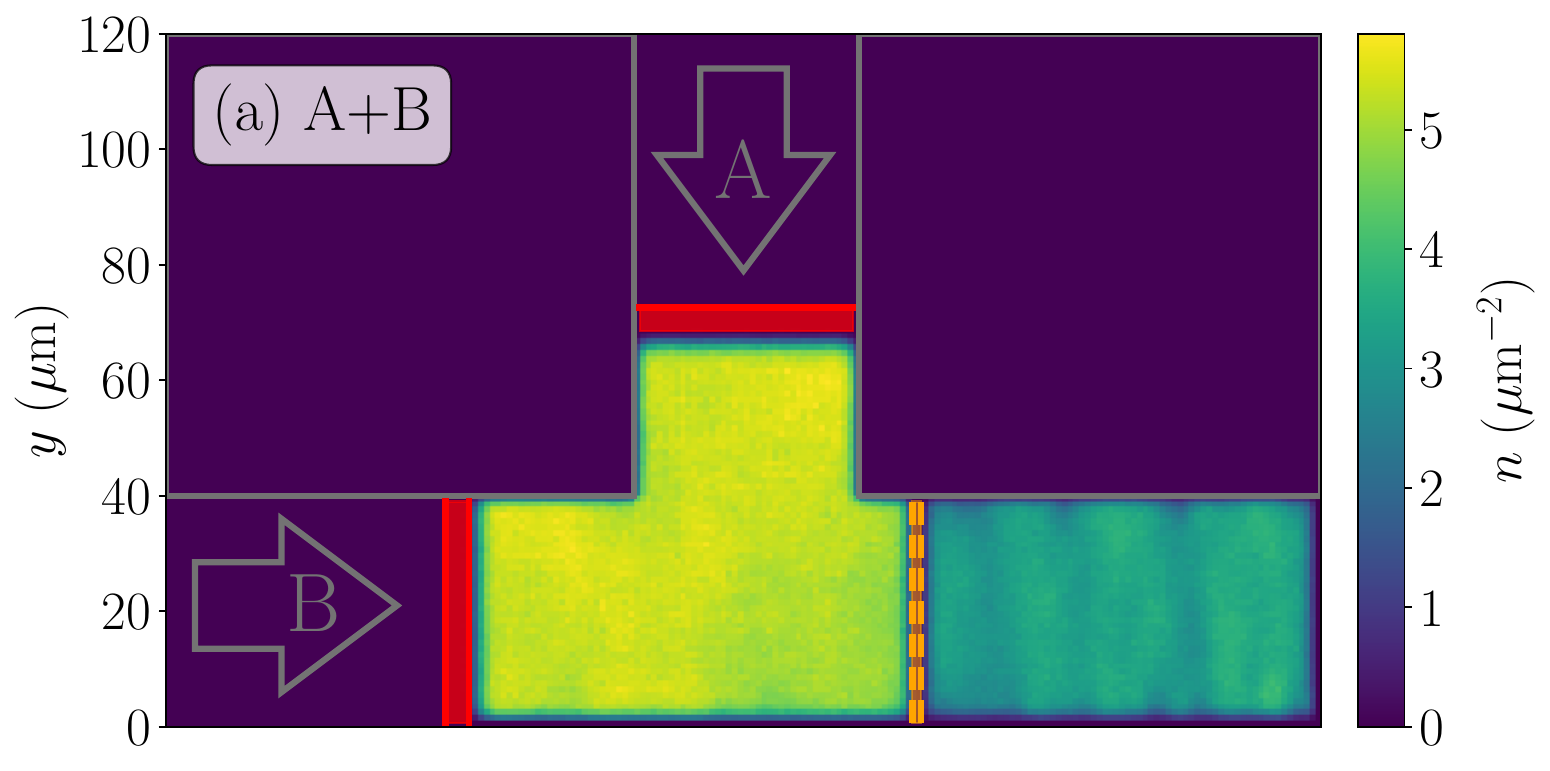}
    \end{minipage}
    \hfill 
    \begin{minipage}{\columnwidth}
        \centering
        \includegraphics[width=\textwidth]{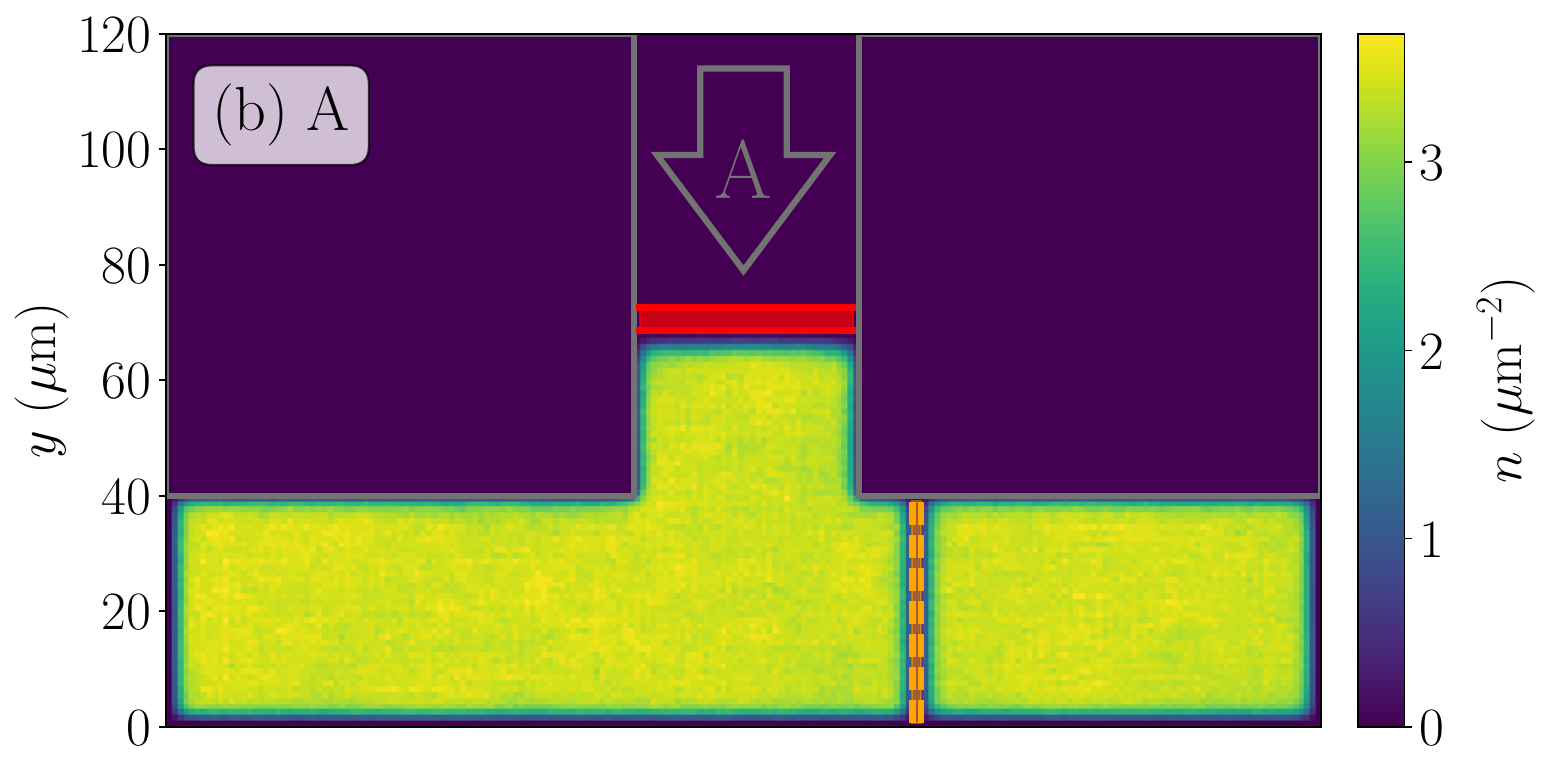}
    \end{minipage} 
    \hfill 
    \begin{minipage}{\columnwidth}
        \centering
        \includegraphics[width=\textwidth]{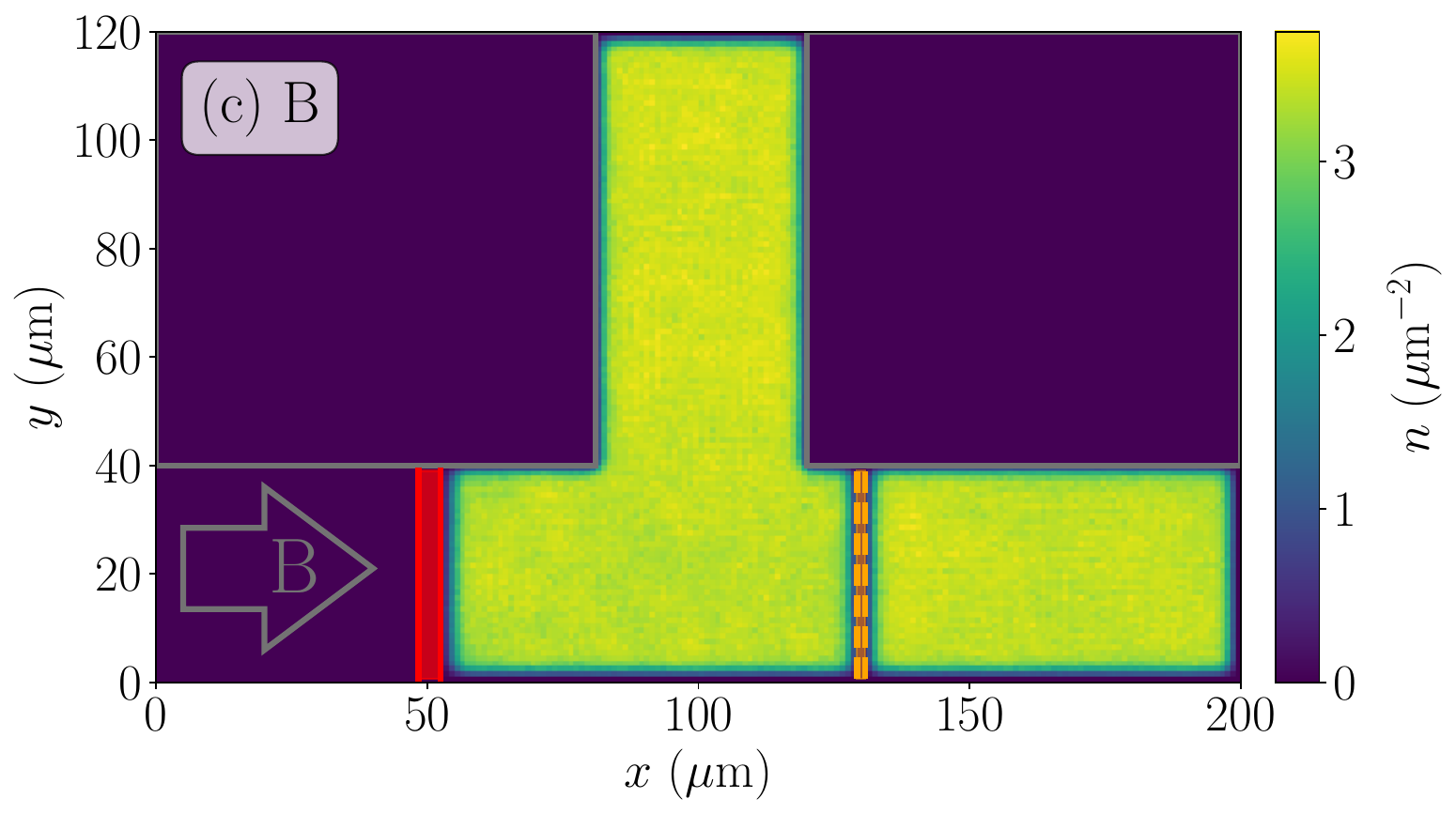}
    \end{minipage}

    \vspace{0.5em}

    \hspace{-25em} 
    \text{(d)} \\[0.4em]
    
    \begin{tikzpicture}[scale=1.04]
      \draw[->] (0,0) -- (7,0) node[right] {t (ms)};
    
      \foreach \x/\y in {0/0, 1/100, 1.5/150, 6.5/650}
      {
        \draw (\x,0.1) -- (\x,-0.1) node[below] {\y};
      }

       \node at (0.5,0.8) [align=center, text width=1.5cm] {Ramp up ($\nearrow$) barriers};
      \node at (4,0.7) [align=center, text width=5.5cm] {Move barriers A/B at\\ constant velocity $v_0$};
      \end{tikzpicture}
    
    \caption{ 
    Dynamical regimes of an atomtronic 2-input AND gate. 
    (a) Density distribution $n(x, y)$, averaged over the initial ensemble, is shown at $t_{\text{end}}=650 \text{ ms}$ after the constant movement of two mobile barriers, labeled A and B (marked in red), with a velocity of $v_0= 0.1 \text{ mm/s}$. 
    These barriers move from the edge of the condensate to their final positions. 
    Each mobile barrier has a height of $\tV_{0, \mm}=3$ and a width of $\tsigma_\mm = 8$. 
    The static barrier, marked in orange, has a height of $\tV_{0, \ms}=2.5$ and width of $\tsigma_\ms = 3$. 
    (b) The same parameter values are used to display the density distribution when only the vertical barrier is moved. 
    (c) $n(x, y)$ is shown for the case of horizontal barrier movement. 
    (d) The barrier protocol. All Gaussian barriers are linearly ramped up over $100$ ms, followed by a $50\,\text{ms}$ waiting period. After this, the barriers move at a constant speed for $500$ ms. Finally, we calculate the density imbalance using densities across the static barrier within the region $ 80\,\mu\text{m} < x < 180\,\mu\text{m}$ and $ 0\,\mu\text{m} < y < 40\,\mu\text{m}$. }
    \label{fig1}
\end{figure}

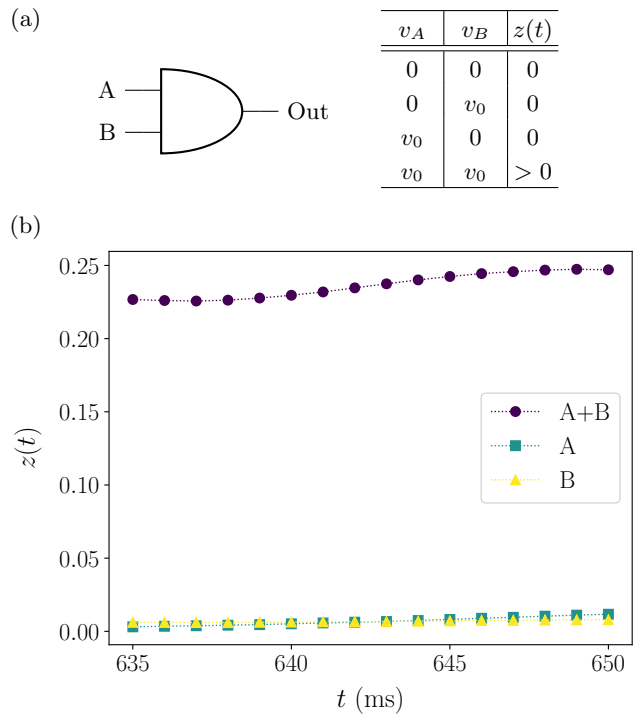
\begin{figure}[h!]
\hspace{-25em} 
    \text{(a)} \\[-1em]
  \begin{minipage}[c]{0.45\linewidth}  
    \centering
    \begin{circuitikz}[scale=1]
      \draw
      (0,0) node[and port] (myand) {}
      (myand.in 1) -- ++(-0.25,0.0) node[left] {A}
      (myand.in 2) -- ++(-0.25,-0.0) node[left] {B}
      (myand.out) -- ++(0.25,0) node[right] {Out};
    \end{circuitikz}
  \end{minipage}%
  \begin{minipage}[c]{0.35\linewidth}
    \centering
    \begin{tabular}{c|c|c}
      \hline
      \,\,\,$v_A$\,\,\, & \,\,\,$v_B$\,\,\, & $z(t)$ \\
      \hline \hline
      $0$ & $0$ & $0$ \\
      $0$  & $v_0$ &  $0$ \\
      $v_0$ & $0$ & $0$ \\
      $v_0$ & $v_0$ & $> 0$ \\
      \hline
    \end{tabular}
    \vspace{1em} 
    \end{minipage}
    \begin{minipage}[c]{1.\linewidth}
    \hspace{-25em} 
    \text{(b)} \\[-0em]
    \includegraphics[width=0.98\columnwidth]{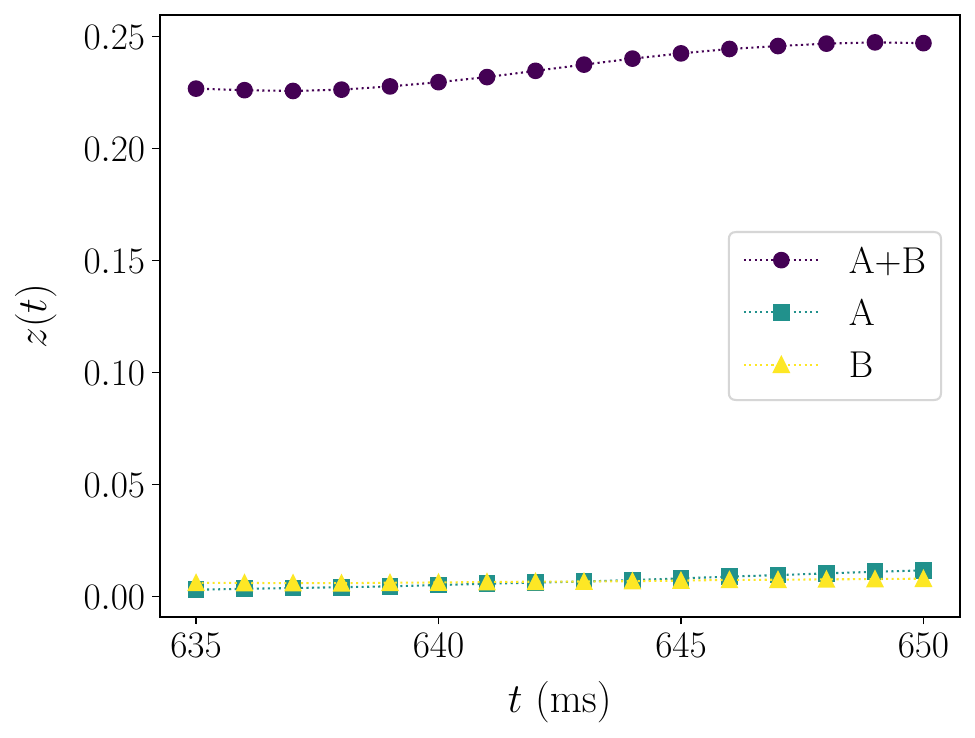}    \vspace{-1em}
    
    \caption{
       Implementation of an atomtronic AND gate. 
       (a) Sketch of a standard AND gate is shown alongside the corresponding truth table for the atomtronic AND gate. 
        $v_A$ and $v_B$ represent the velocities of the individual barrier movements, while $z(t)$ denotes the density imbalance across the static barrier.  
       (b) Time evolution of the imbalance $z(t) $ is shown after the completion of the barrier movements; see text.
        When only the barrier A is moved or only the barrier B is moved, the imbalance remains relatively small. However, a significant imbalance is observed when both barriers A and B are moved together. }     
    \label{fig2}
\end{minipage}
\end{figure}

\section{\label{sec:2InputAndGate}Two-Input AND Gate}

In this section, we demonstrate the realization of an atomtronic AND gate in a 2D homogeneous T-shaped BEC, as shown in Fig. \ref{fig1}.
The system consists of a closed design with dimensions $L_x \times L_y = 200 \times 120 \,\mu\text{m}^2$. 
The AND gate acts as a circuit element exhibiting transistor-like behavior, 
which we implement by introducing three Gaussian barriers in each arm of the reversed T-shaped BEC.  
Two of these barriers, labeled as A and B, are mobile and positioned at the edges of the condensate at $x_{0, \mm}=0 \,\mu\text{m}$ and $y_{0, \mm}=120\,\mu\text{m}$. 
In our proposed operation each of these barriers is moved either at zero velocity or at a velocity $v_0$. This binary choice encodes the input of the logic gate. These barriers are sufficiently high to not allow tunneling across them. Therefore, they act as classical pistons and induce currents as they push the atomic superfluid. The third, static barrier in the lower right part of the inverse-T shape, has a barrier height that allows for tunneling across the barrier. Therefore it acts as a Josephson junction on the superfluid dynamics. In particular, it displays a critical current behavior. If the current across the junction is smaller than it, the density imbalance across it remains small. However, if the current is larger than the critical current,  significant density imbalance builds up, corresponding to the resistive regime.
This was used to determine the current-chemical potential relationship in atomic JJs \cite{Kwon2020, Pace2021}, 
where the finite chemical potential regime occurs via vortex emission \cite{Burchianti2018, Xhani2020, singh2023shapiro}.
For barriers A and B, we use $\tV_{0, \mm}=3$, $\tsigma_\mm=8$ and equal displacement of $\Delta x\, = \Delta y = 50\,\mu \text{m}$. 
The velocity is chosen to be $v=v_0 = 0.1$ mm/s. 
This velocity induces a current at the static barrier that is below the critical current, if only one of the mobile barriers is moved. We refer to the barrier velocity that creates the critical current as $v_c$, so we have $v_0 < v_c$.
The static barrier is positioned at $x_{0, \ms} = 130\,\mu\text{m}$ in the right horizontal arm, with $\tV_{0, \ms}=2.5$ and $\tsigma_\ms=3$.

The key mechanism to create an AND gate, or a transistor-like behavior, is that if both barriers A and B are moved at $v_0$, the resulting total current exceeds the critical current, producing a density imbalance at the static barrier, see Fig. \ref{fig1}(a).   
In contrast, when either A or B is moved independently at $v_0$, the junction remains in the subcritical regime, allowing supercurrent tunneling, and no density imbalance appears at the static barrier, see Figs. \ref{fig1}(b) and \ref{fig1}(c). To quantify these observations, we calculate the imbalance across the static barrier as 
\begin{equation} \label{eq:imbalance}
z(t) =  \frac{n_\text{left}(t) - n_\text{right}(t)}{n},
\end{equation}
where $n_\text{left}(t)$ and $n_\text{right}(t)$ represent the densities in the left and right reservoirs across the static barrier, here determined within the region $ 80\,\mu\text{m} < x < 180\,\mu\text{m}$ and $ 0\,\mu\text{m} < y < 40\,\mu\text{m}$. 
In Fig. \ref{fig2}(a) we show the time evolution of $z(t)$ after the movement of the barriers for three different input configurations. 
When either A or B moves alone, no imbalance is observed. 
However, when both A and B move together, a nonzero imbalance is created. 
These results are summarized in the table in Fig. \ref{fig2}(a), 
which corresponds to the logic table of an AND gate.

\begin{figure*}[hbt]
    \centering
        \includegraphics[ height=7.4cm]{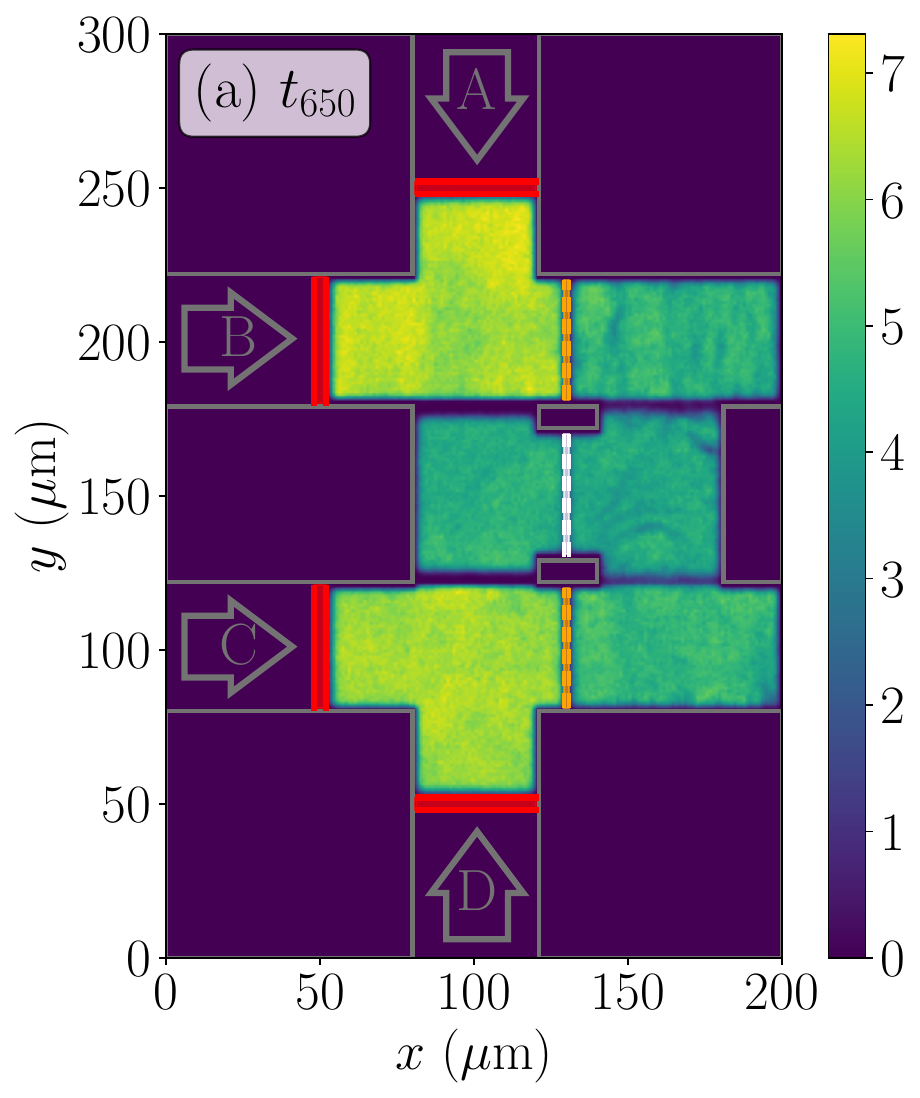}\hspace{-0em}
        \includegraphics[height=7.3cm]{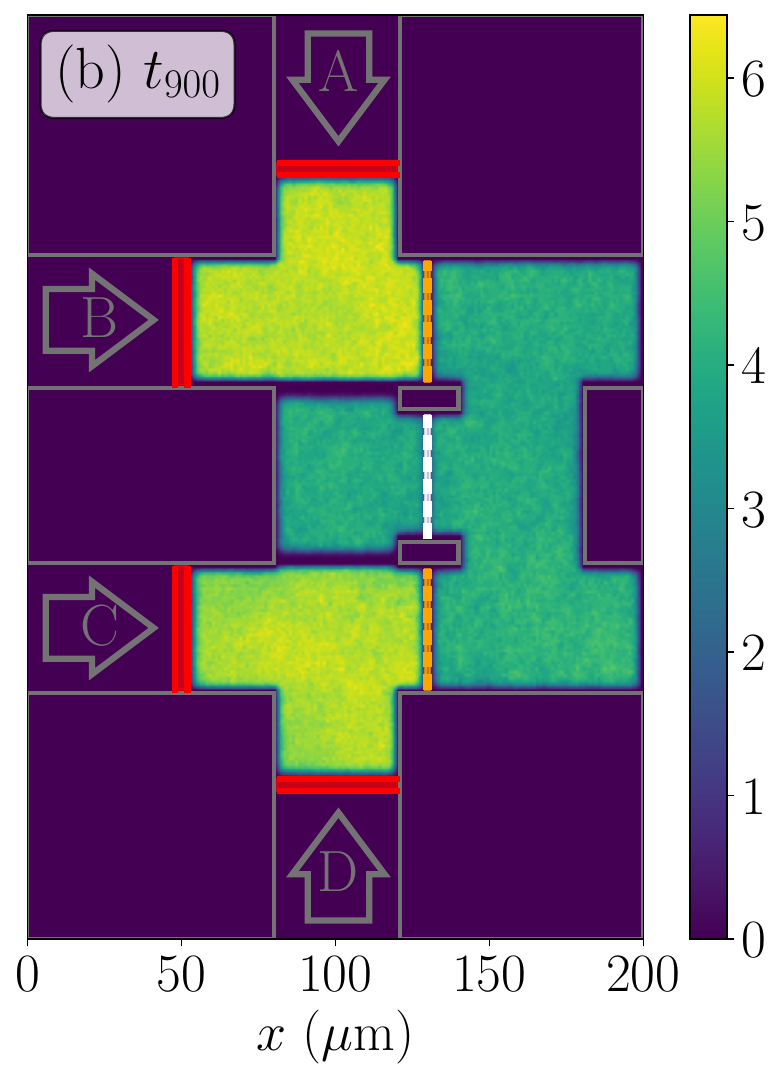}\hspace{-0em}
        \includegraphics[height=7.3cm]{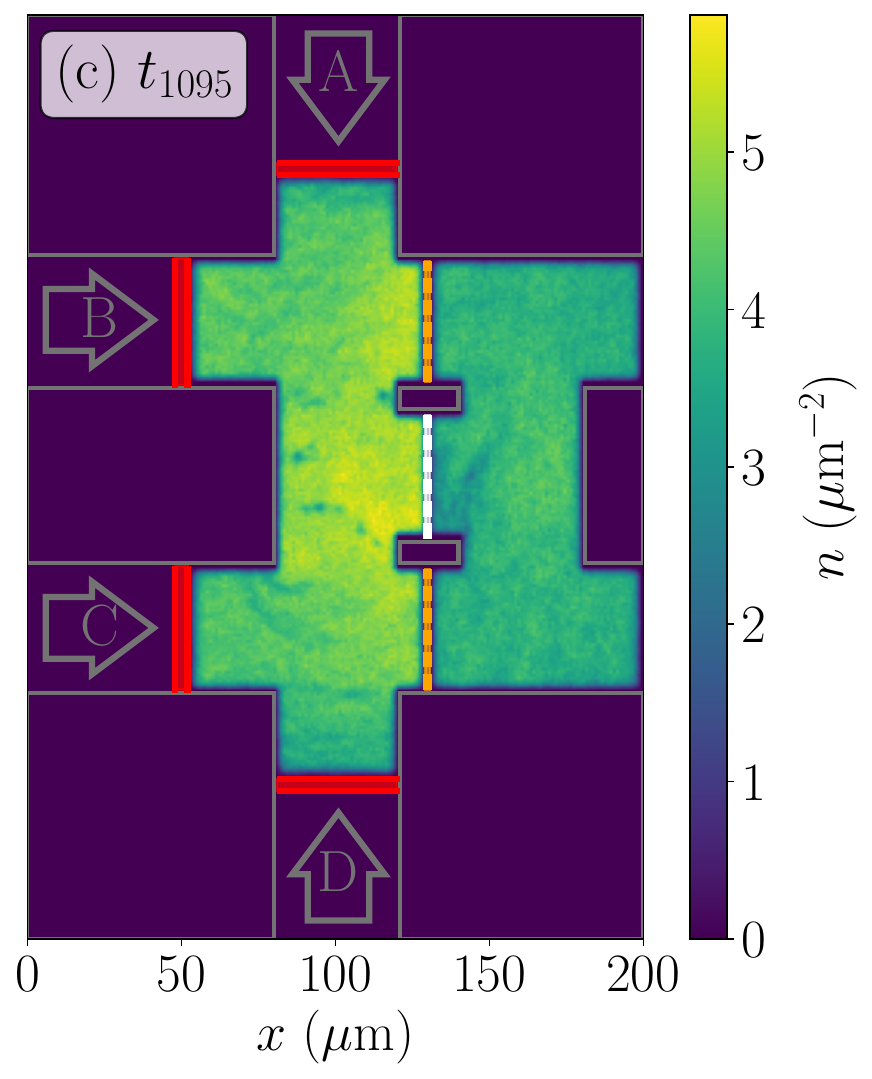}
        
        \vspace{1em}

        \hspace{-53em} 
    \text{(d)} \\[-0.5em]
        
        \begin{tikzpicture}[scale=1.34]
          \draw[->] (0,0) -- (11,0) node[right] {t (ms)};
        
          \foreach \x/\y in {0/0, 1/100, 1.5/150, 6.5/650, 8.5/850, 9/900, 10.5/1050}
          {
            \draw (\x,0.1) -- (\x,-0.1) node[below] {\y};
          }
        
           \node at (0.5,0.5) [align=center, text width=2cm] {Ramp up ($\nearrow$) barriers};
          \node at (4,0.5) [align=center, text width=5.5cm] {Move barriers A, B, C, D \\at constant velocity $v_0$};
          \node at (7.5,0.5) [align=center, text width=3cm] {Open ($\searrow$) right output gates};
          \node at (9.75,0.5) [align=center, text width=2.2cm] {Open ($\searrow$) left output gates};
          \end{tikzpicture}
          
    \caption{
    Implementation of an atomtronic circuit to create a 4-input AND gate. 
    (a) We show the density distribution $n(x,y)$ at time $t=650\text{ ms}$, for the case of moving all four mobile barriers (A, B, C and D) simultaneously at a constant velocity of $v_0= 0.1 \text{ mm/s}$ from the edges of the condensate to their final positions. Each two-input AND gate consists of two mobile barriers with heights $\tV_{0, \mm}=3$ and widths $\tsigma_\mm= 8$, and one static barrier with a height $\tV_{0, \ms}=2.5$ and width $\tsigma_\ms= 3$ (orange dashed lines).  
    Four Gaussian barriers, each with a height of $\tV_{0, \ms}=5$ and a width of $\tsigma_\ms = 3$, separate the two-input AND gates from the central part,  functioning as left and right output gates. A static barrier with a height of $\tV_{0, \ms}=2$ and width $\tsigma_\ms= 2$ (white dashed lines) is placed in the central part, acting as the output channel. 
    (b)  At $t=900\text{ ms}$, $n(x,y)$ is shown after the right output gates are opened. (c) At $t=1095\text{ ms}$,  $n(x,y)$ is displayed after the left output gates are opened. 
    (d) Timeline of the barrier protocol. Following the AND-gate procedure shown in Fig.  \ref{fig1}(d), the left and right output gates are linearly ramp down to analyze the resulting density buildup within the area of $ 80\,\mu\text{m} < x < 180\,\mu\text{m}$ and $130\,\mu\text{m} < y < 170\,\mu\text{m}$.  }
    \label{fig3}
\end{figure*}
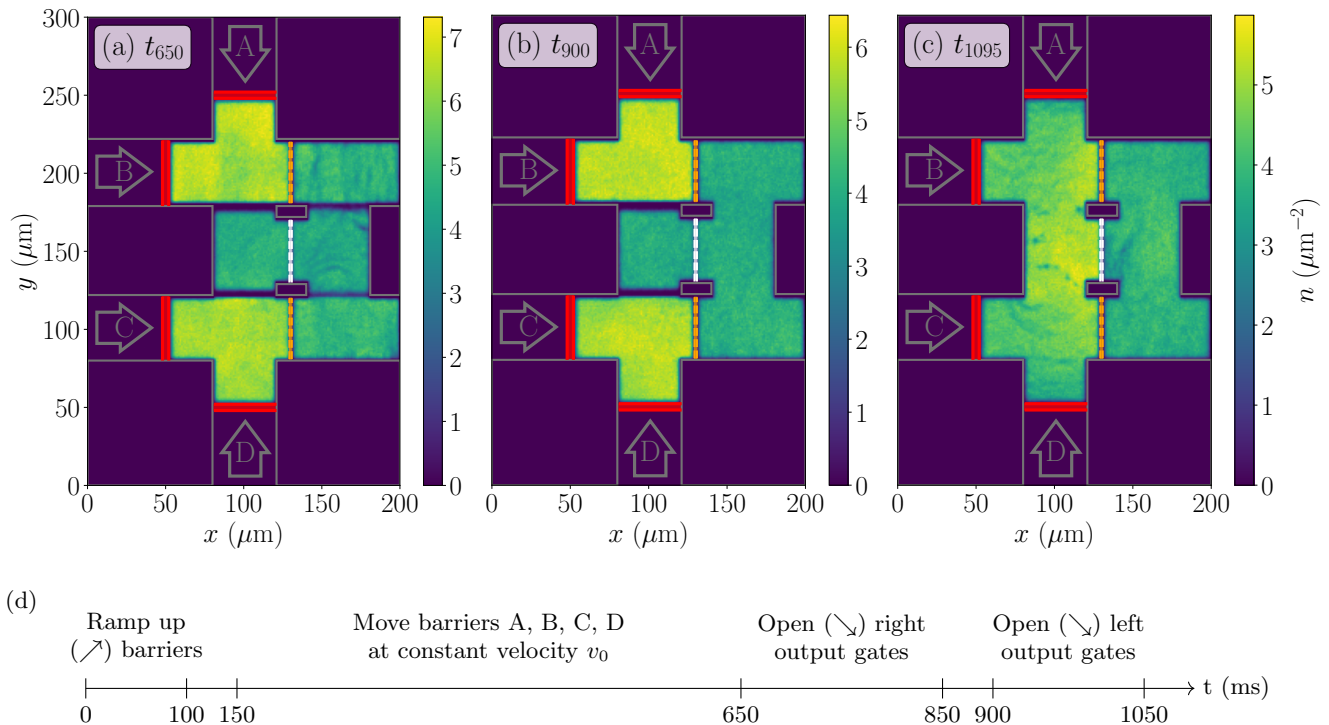

\section{\label{sec:4InputAndGate} Circuit of AND gates}
Having implemented a logical AND gate in Sec. \ref{sec:2InputAndGate}, we extend this concept to create a more complex atomtronic circuit. Specifically, we expand the 2-input AND gate to a circuit of AND gates to create a 4-input AND gate, as shown in Fig. \ref{fig3}. 
For this, we connect two T-shaped structures, each providing two inputs, within an H-shaped configuration that serves as the designated output. 
The H-shaped connector couples to the regions located to the left and right of the static barriers that were previously analyzed for the density imbalance in the 2-input AND gate. 
For both T-shaped circuits we use the same parameters as in the previous section. 
The barrier protocol for the 4-input AND gate is shown in Fig. \ref{fig3}(d).

In this setup, a static barrier is placed within the H-shaped structure at $x_{0, \ms}=130 \,\mu \mm$, with parameters $\tV_{0, \ms}=2$ and $\tsigma_\ms = 2$. 
This new barrier is slightly weaker than the two static barriers used in the top and bottom parts of the setup. The areas adjacent to this new barrier now constitute the output of the 4-input AND gate circuit, specifically, whether a density imbalance builds up between them, or not. 
Furthermore, we introduce four Gaussian barriers, two left and two right output gates, at $y=120\,\mu\text{m}$ and $y=180\,\mu\text{m}$, connecting the top, bottom, and middle regions. 
The parameters for these gates are $\tV_{0, \ms}=5$ and $\tsigma_\ms = 3$.  
These barriers prevent particles from tunneling through prematurely, ensuring that the initial AND-gate protocols complete without interference, and that the generated output remains unaltered.
At $t=650$ ms, after both 2-input AND gates in the top and bottom area have completed their operations, the mobile barriers are in their final positions. 
We then linearly ramp down the right output gates from $\tV_{0, \ms}=5$ to $0$ over a period of $200 \,$ms. After this step, a waiting period of $50$ ms allows the system to stabilize. 
Finally, we gradually open the left output gates over $150\,$ms, followed by another $50$ ms waiting time to allow the system to settle. 
In Fig. \ref{fig3}, this process is depicted for the case of moving all input barriers, 
resulting in a nonzero imbalance across the new static barrier. 
For comparison, an example resulting in zero imbalance is discussed in Appendix \ref{sec:appA}.

\begin{figure}[t!]
\hspace{-25em} 
    \text{(a)} \\[-2em]
    \begin{minipage}[c]{1.\linewidth}
    \begin{circuitikz}
    \centering
      \draw
      (0,0) node[and port] (myand1) {}
      (myand1.in 1) -- ++(-0.5,0.0) node[left] {A}
      (myand1.in 2) -- ++(-0.5,-0.0) node[left] {B}

      (0,-1.5) node[and port] (myand2) {}
      (myand2.in 1) -- ++(-0.5,0.0) node[left] {C}
      (myand2.in 2) -- ++(-0.5,-0.0) node[left] {D}

      (2.5,-0.75) node[and port] (myand3) {}
      (myand1.out) -- (myand3.in 1)
      (myand2.out) -- (myand3.in 2)

      (myand3.out) -- ++(0.3,0) node[right] {Out};
    \end{circuitikz}
\end{minipage}
    \begin{minipage}[c]{1.\linewidth}
    \hspace{-25em} 
    \text{(b)} \\[-0em]
    \includegraphics[width=0.98\columnwidth]{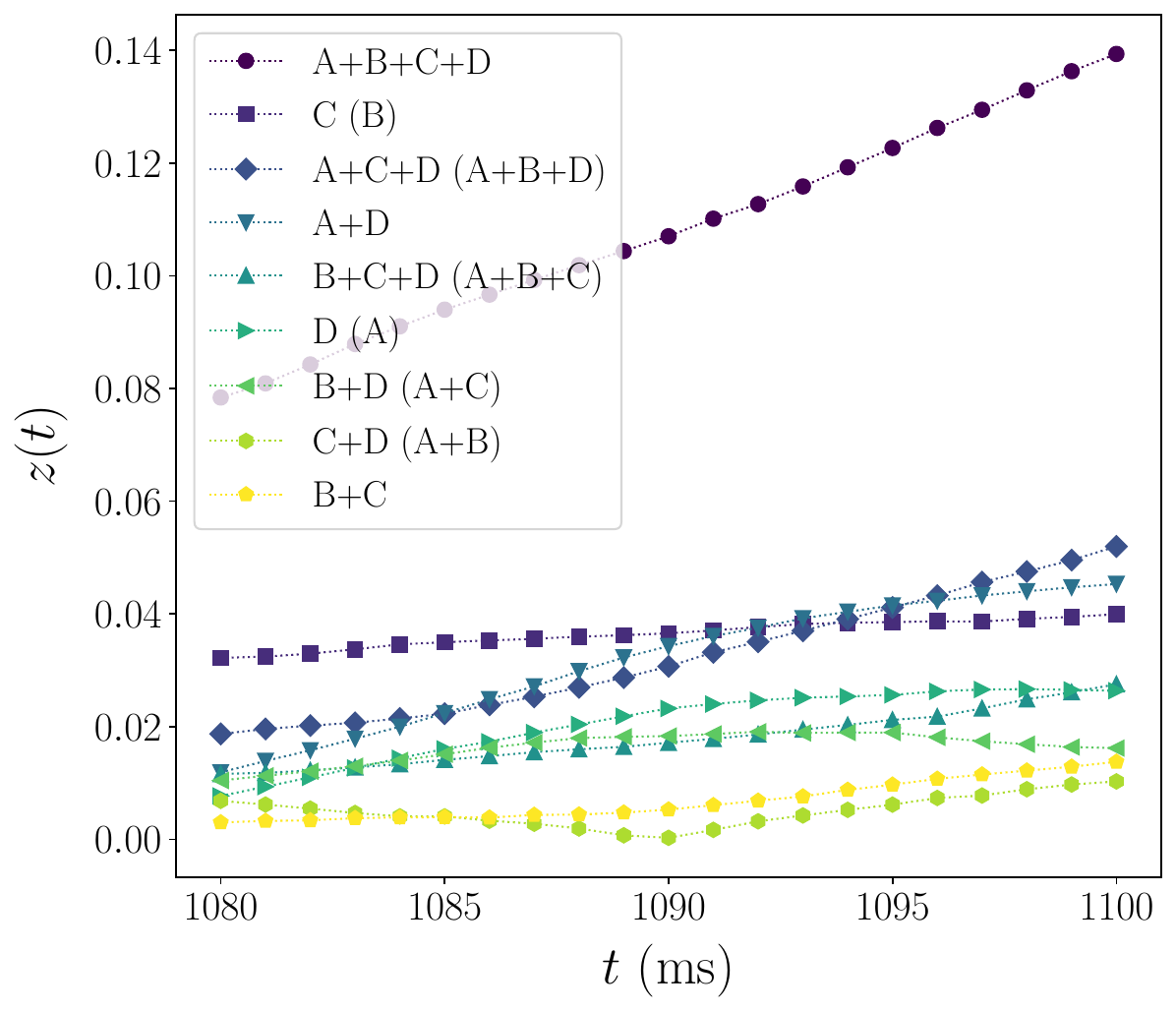} 
    \caption{Results of the logical atomtronic 4-input AND gate. 
    (a) Sketch of a 4-input AND gate. (b) Time evolution of the density imbalance $z(t)$ is shown for all relevant cases among the 16 possible combinations of the 4 mobile input barriers (A, B, C and D). In brackets are the cases that behave similarly due to the symmetry of the setup.}
    \label{fig4}
    \end{minipage}
\end{figure}
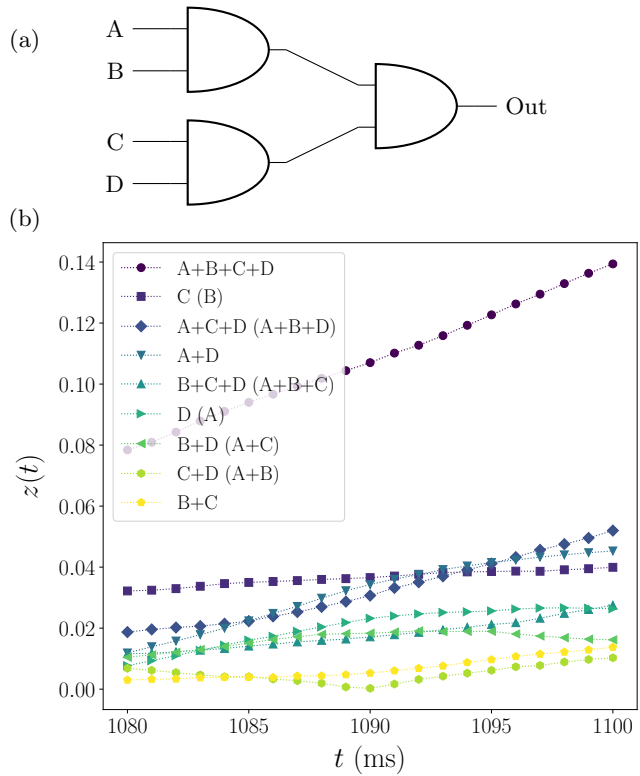

We assign the circuit output based on the density imbalance created across the output barrier within the area of $ 80\,\mu\text{m} < x < 180\,\mu\text{m}$ and $130\,\mu\text{m} < y < 170\,\mu\text{m}$. 
We calculate the imbalance $z$ according to Eq. \ref{eq:imbalance}. 
In Fig. \ref{fig4}(b), we show the time evolution of $z(t)$ for all cases, up to symmetric combinations.
As a key result we observe that only in the case where the barriers A, B, C, and D are moved simultaneously, a clear, nonzero density imbalance emerges, corresponding to a logical $1$. 
All other cases generate near-zero imbalance corresponding to a logical $0$.

\begin{figure}[t]
    \hspace{-25em} 
    \text{(a)} \\[-2em]
  \begin{minipage}[c]{0.6\linewidth}  
    \centering \hspace{0em}
    \begin{circuitikz}
        \draw
        (3,0) node[not port] (mynot) {}
        (mynot.in) -- ++(-0.35,0) node[left] {A}
        (mynot.out) -- ++(0.35,0) node[right] {Out};
    \end{circuitikz}
  \end{minipage}
  \begin{minipage}[c]{0.3\linewidth}
    \centering \vspace{1em} \hspace{-4em}
    \begin{tabular}{c|c}
      \hline
       \,\,\,$z(0)$\,\,\, & $z(t_\text{end})$ \\
      \hline \hline
      $0$  & $>0$ \\
      $>0$ & $0$ \\
      \hline
    \end{tabular}\vspace{1em}
  \end{minipage} 
  
\begin{minipage}[c]{1.\linewidth}
    \hspace{-25em} 
    \text{(b)} \\[-0em]
    \includegraphics[width=0.9\columnwidth]{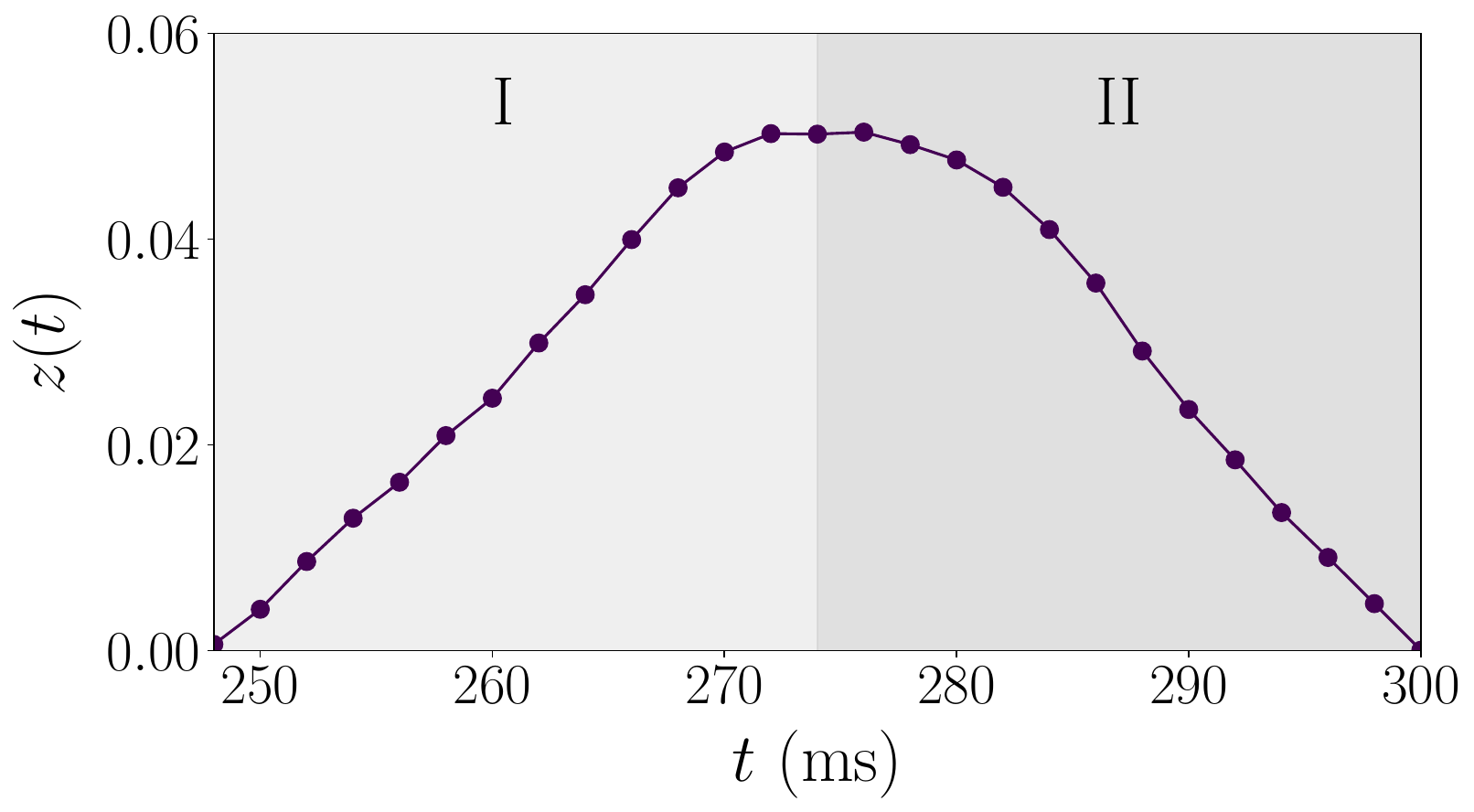} 
    \vspace{-0.5em}\\ \hspace{-25em}
    \text{(c)} \\[0.4em]
\begin{tikzpicture}[scale=1.8]
      \draw[->] (0,0) -- (3.5,0) node[right] {t (ms)};
    
      \foreach \x/\y in {0/0, 1/100, 1.5/150, 3.0/300}
      {
        \draw (\x,0.1) -- (\x,-0.1) node[below] {\y};
      }

       \node at (0.5,0.7) [align=center, text width=1.5cm] {Ramp up ($\nearrow$) barriers};
      \node at (1.5,0.68) [align=center, text width=1.5cm] {Phase Imprint $\phi_0$};
      \end{tikzpicture}
  \vspace{1em}
    \caption{
    (a) NOT gate and the corresponding truth table for the atomtronic NOT gate. $z(0)$ represents the initial imbalance, and $z(t_\text{end})$ is the final imbalance. 
    (b) Time evolution of the imbalance $z(t)$ is shown following a phase imprint of $\phi_0 = 0.25\, \pi$ on the left subsystem. $z(t)$ undergoes half a Josephson oscillation (JO), with regions I and II (each representing a quarter of the JO) illustrating the imbalance dynamics for the two NOT-gate operations.(c) Barrier protocol. A static barrier is linearly ramped up at the center of the cloud over $100$ ms, followed by a $50\,\text{ms}$ waiting period. At $t=150$, we imprint a phase $\phi_0 = 0.25\, \pi$ on the left subsystem. }
    \label{fig:NOT/NAND-gate Symbol + Table}
\end{minipage}

\end{figure}
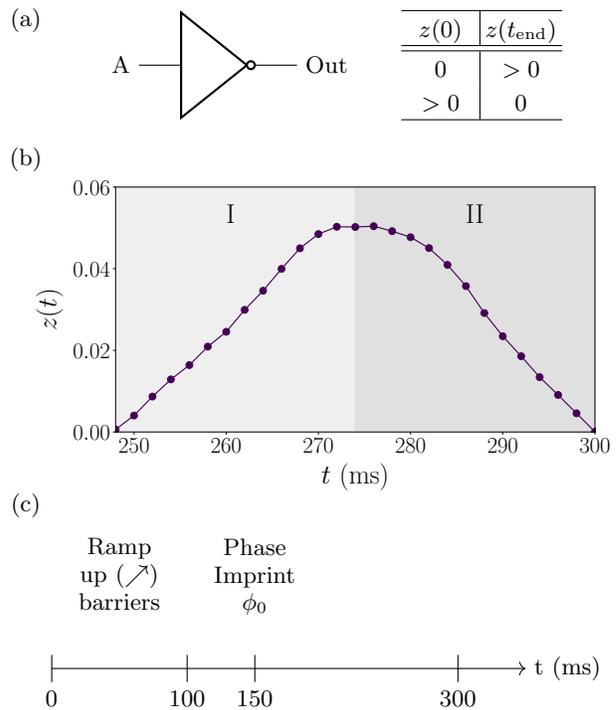

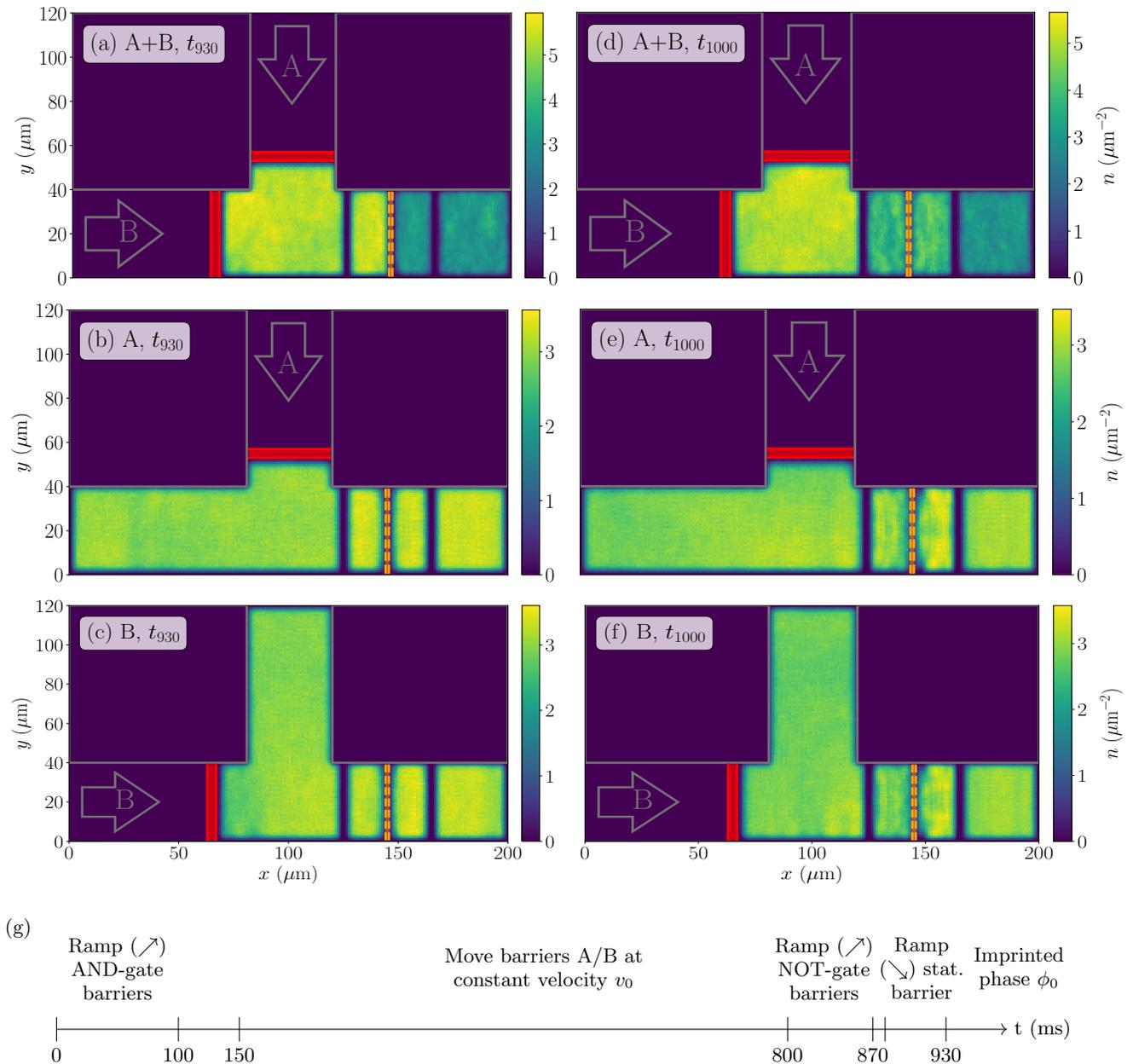
\begin{figure*}[t]
    
        \centering
        \includegraphics[width=\columnwidth, height=4.6cm]{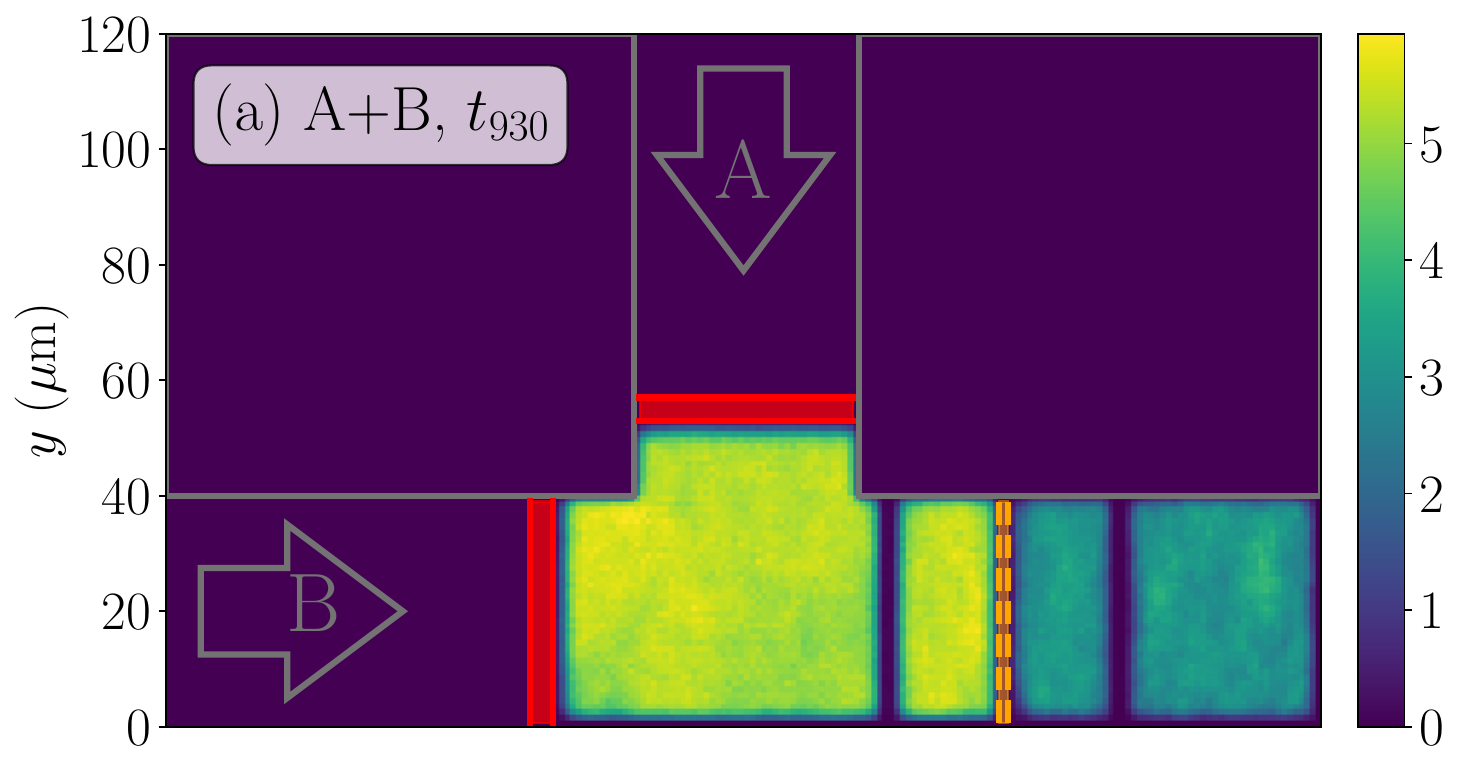}\hspace{-0em}
        \includegraphics[width=\columnwidth, height=4.5cm]{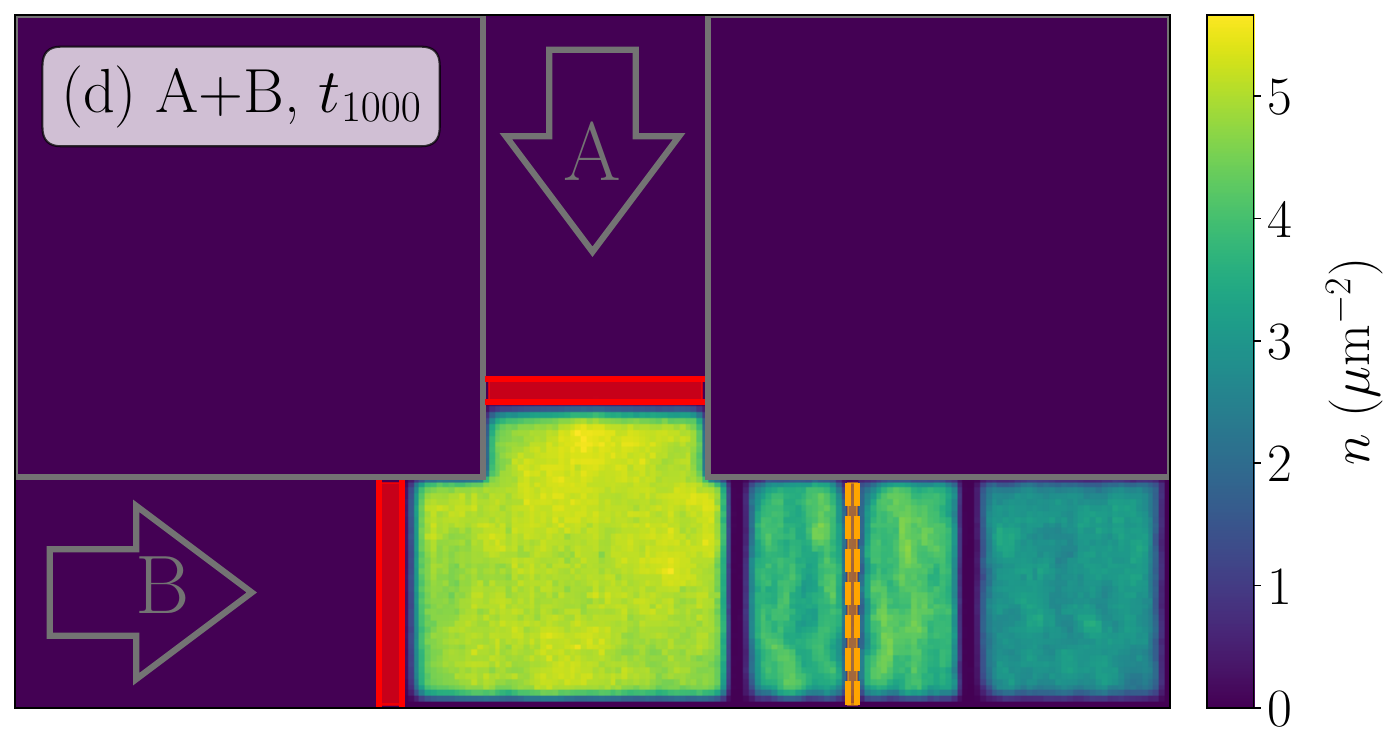}

        \includegraphics[width=\columnwidth, height=4.6cm]{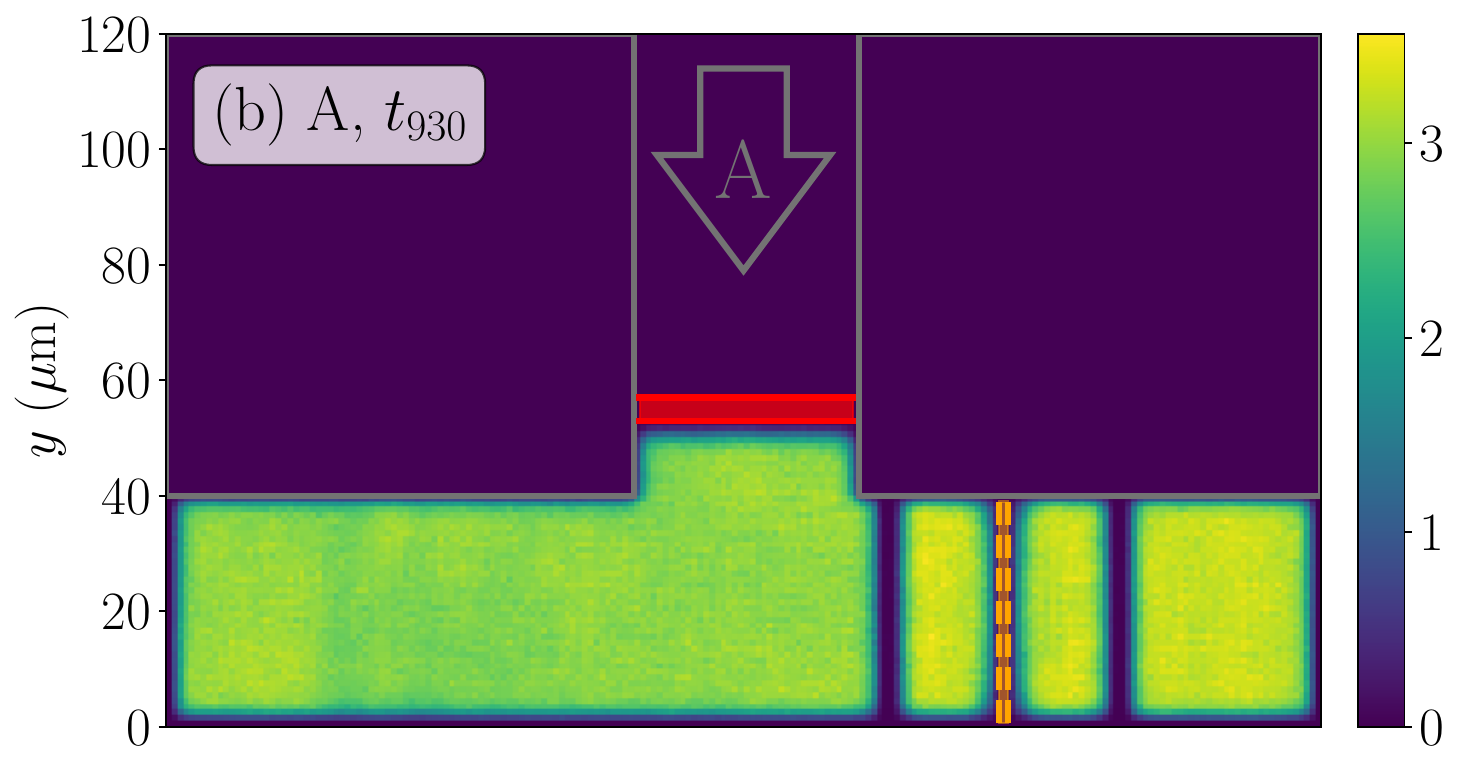}
        \hspace{-0em}
        \includegraphics[width=\columnwidth, height=4.5cm]{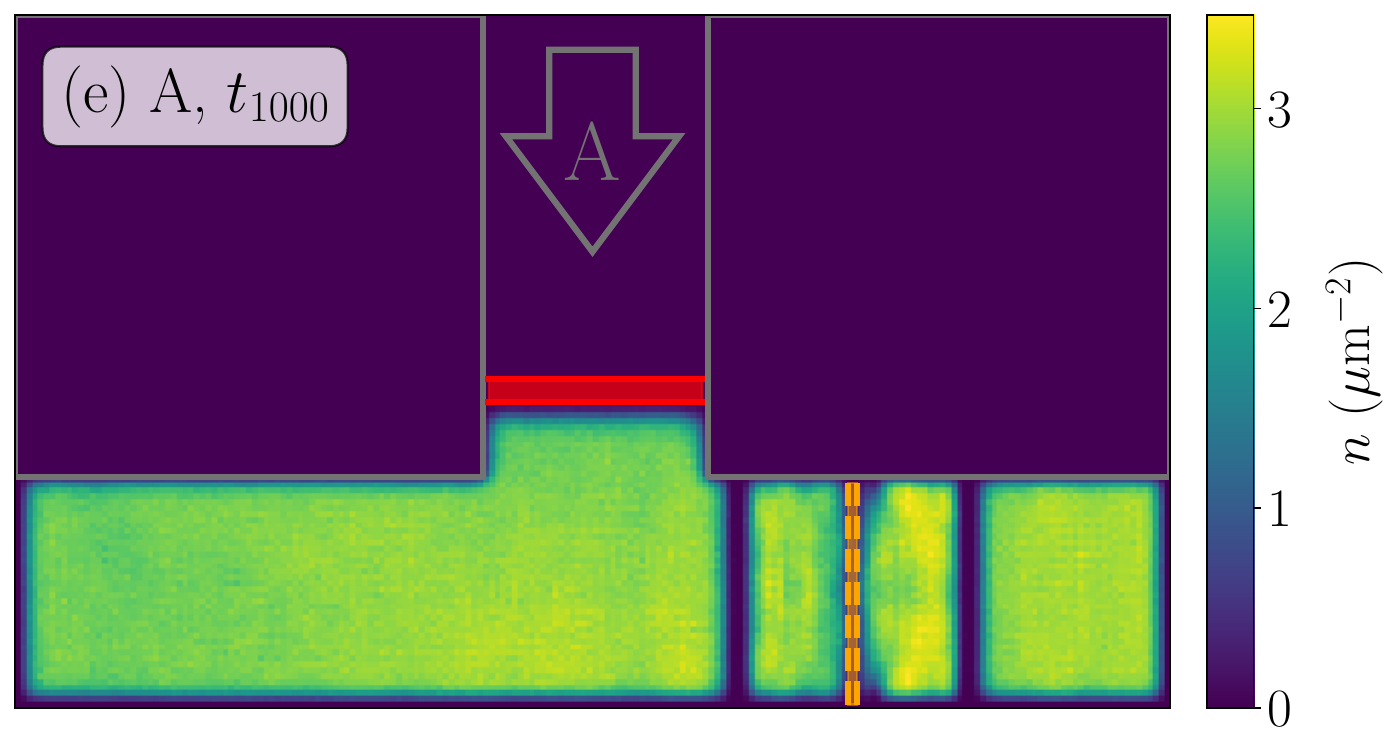}
        
        \includegraphics[width=\columnwidth, height=4.6cm]{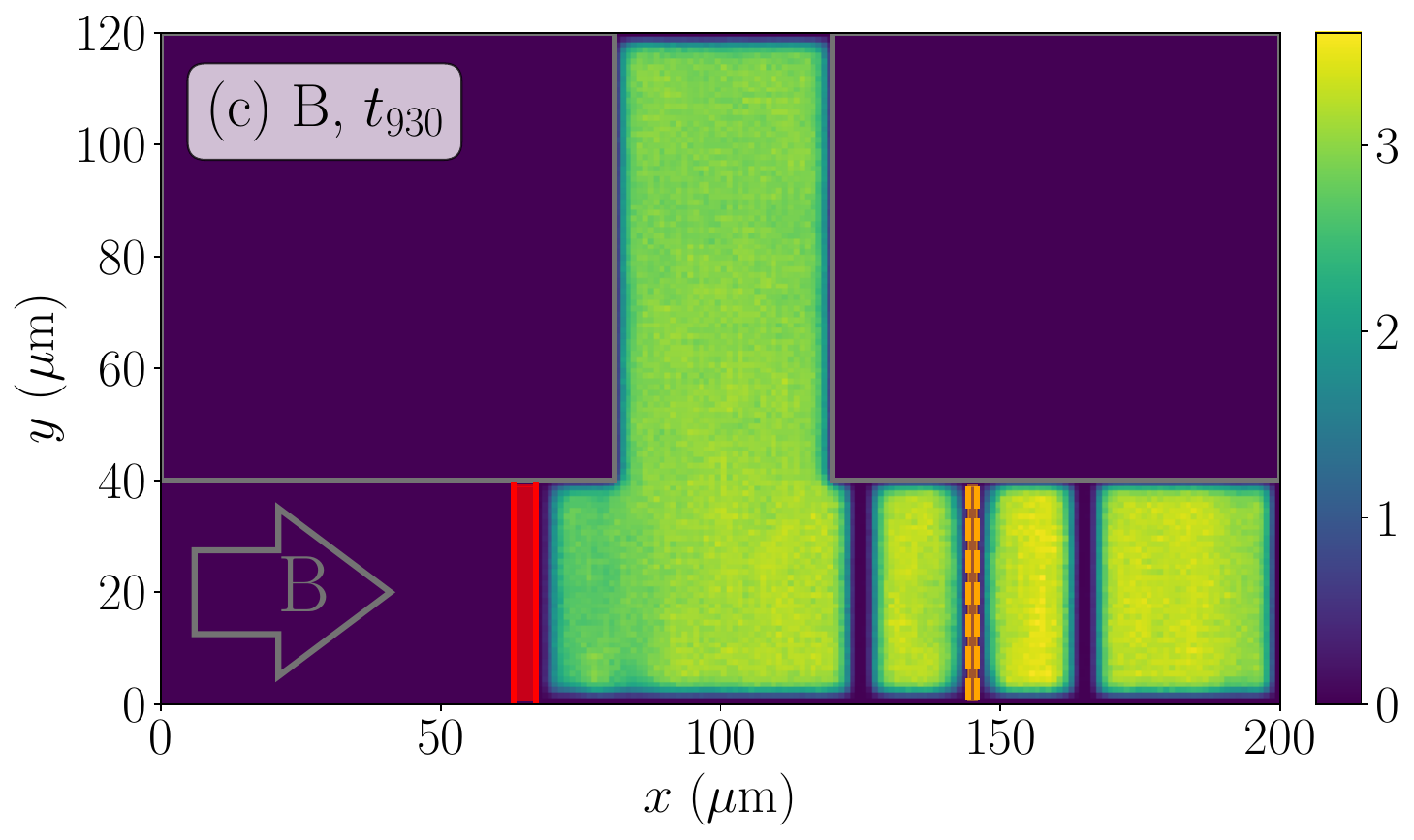}
        \hspace{-0em}
        \includegraphics[width=\columnwidth, height=4.5cm]{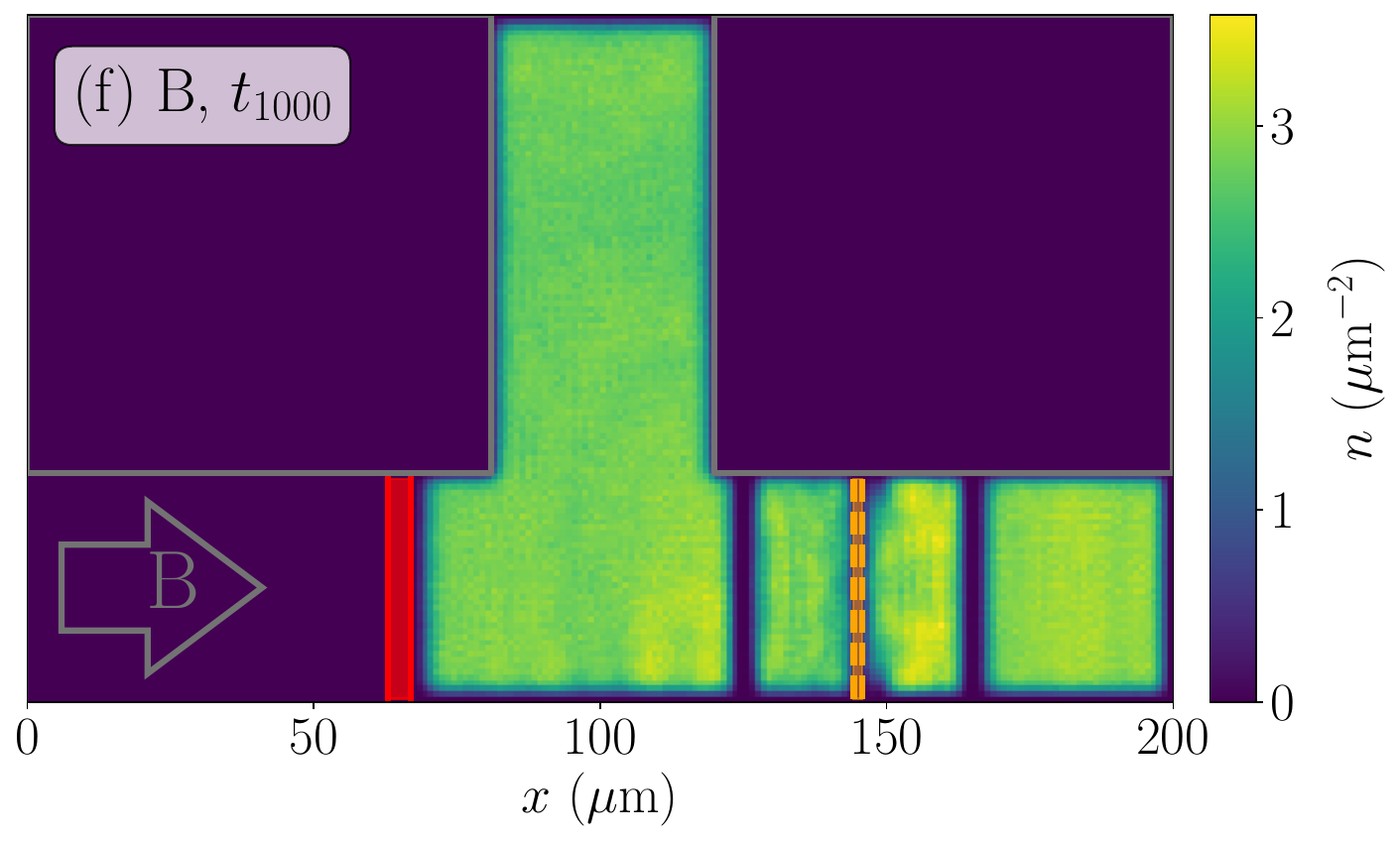}

        \vspace{1em}
        \hspace{-53em} 
    \text{(g)} \\[-0.5em]
        \begin{tikzpicture}[scale=1.9]
          \draw[->] (0,0) -- (7.8,0) node[right] {t (ms)};
        
          \foreach \x/\y in {0/0, 1/100, 1.5/150, 6.0/800, 6.7/870,  7.3/930}
          {
            \draw (\x,0.1) -- (\x,-0.1) node[below] {\y};
          }

           \foreach \x/\y in {6.8/880}
          {
            \draw (\x,0.1) -- (\x,-0.1) ;
          }
        
           \node at (0.5,0.5) [align=center, text width=2cm] {Ramp ($\nearrow$) AND-gate barriers};
          \node at (4,0.5) [align=center, text width=5.5cm] {Move barriers A/B at\\ constant velocity $v_0$};
          \node at (6.3,0.5) [align=center, text width=1.75cm] {Ramp ($\nearrow$) NOT-gate barriers};
          \node at (7.1,0.5) [align=center, text width=1.75cm] {Ramp ($\searrow$) stat. barrier}; 
          \node at (7.9,0.5) [align=center, text width=2cm] {Imprinted phase $\phi_0$}; 
          \end{tikzpicture}

    \caption{
    Dynamical regimes of an atomtronic NAND gate. (a,b,c) Density 
    distribution $n(x,y)$ is shown for three different configurations of the input barriers (A and B).
    (a) Both A and B are moved together. (b) Only A is moved, and (c) only B is moved. 
    Two static barriers, positioned at $x_{0, \ms}=125\,\mu \mm$ and $x_{0, \ms}=165\,\mu \mm$,  define the NOT-gate area, 
    while the static barrier at $x_{0, \ms}=145 \,\mu \mm$ (indicated by orange dashed lines) serves as the output barrier. (d,e,f) The corresponding cases are shown after the phase imprint of $\phi_0 = 0.25 \, \pi$ on the left subsystem within the NOT-gate area. 
    (g) Timeline of the barrier protocol.}  
    \label{fig:NAND}
\end{figure*}

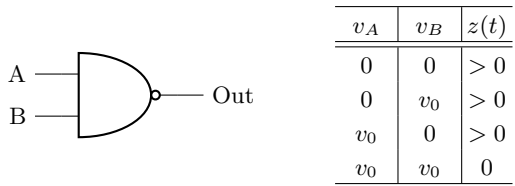
\begin{figure}[h]
    \begin{minipage}[c]{0.45\linewidth}  
    \centering
    \begin{circuitikz}
        \draw
        (6,0) node[nand port] (mynand) {}
        (mynand.in 1) -- ++(-0.35,0) node[left] {A}
        (mynand.in 2) -- ++(-0.35,0) node[left] {B}
        (mynand.out) -- ++(0.35,0) node[right] {Out};
    \end{circuitikz}
  \end{minipage}%
  \begin{minipage}[c]{0.45\linewidth}
    \centering
    \begin{tabular}{c|c|c}
      \hline
      \,\,\,$v_A$\,\,\, & \,\,\,$v_B$\,\,\, & $z(t)$ \\
      \hline \hline
      $0$ & $0$ & $> 0$ \\
      $0$  & $v_0$  & $> 0$ \\
      $v_0$ & $0$ & $> 0$ \\
      $v_0$ & $v_0$ & $ 0$ \\
      \hline
    \end{tabular}
    \end{minipage}
    \caption{Sketch of a standard NAND gate and the corresponding truth table for the atomtronic NAND gate. $v_A$ and $v_B$ denote the velocities of the individual barrier movements. $z(t)$ represents the final imbalance determined after the phase imprint. }
    \label{fig:NOT/NAND-gate Symbol + Table}
\end{figure}

\begin{figure*}[hbt]
    \centering
        \includegraphics[height=7.4cm]{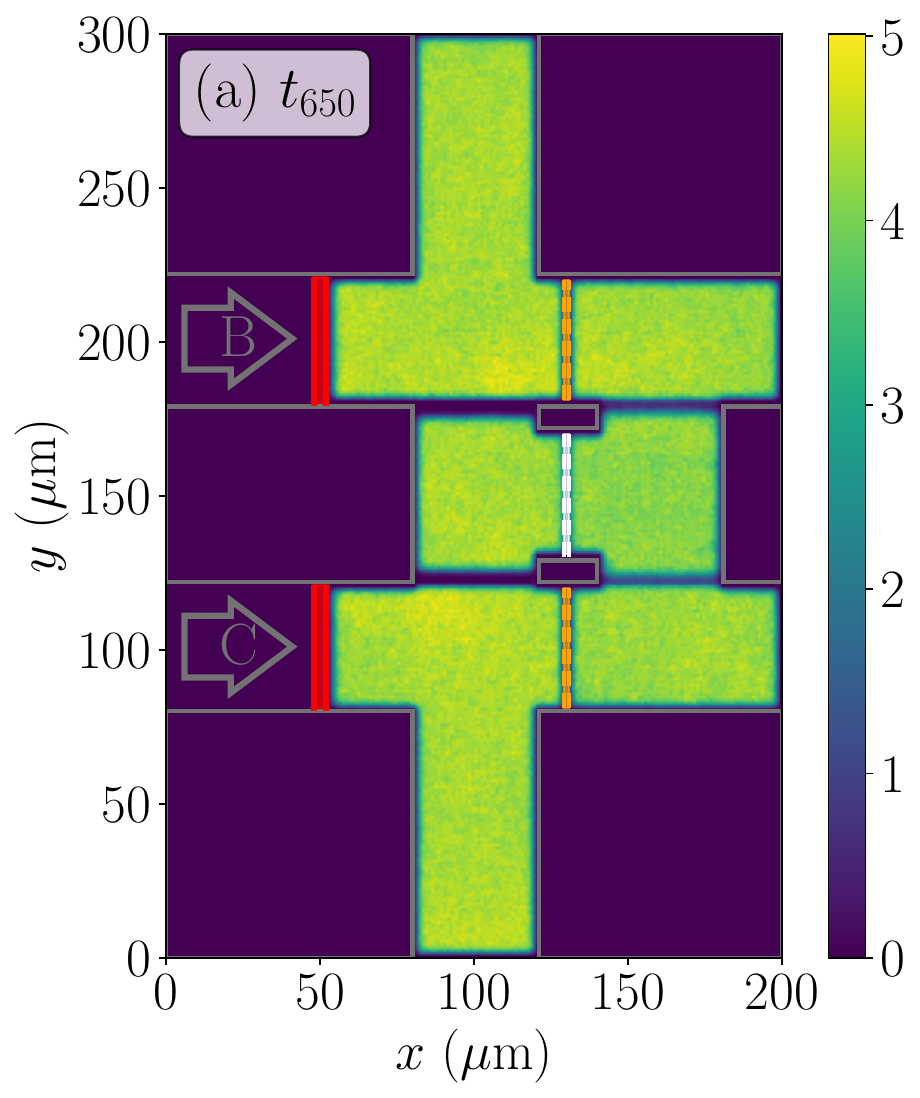}\hspace{-0em}
        \includegraphics[height=7.3cm]{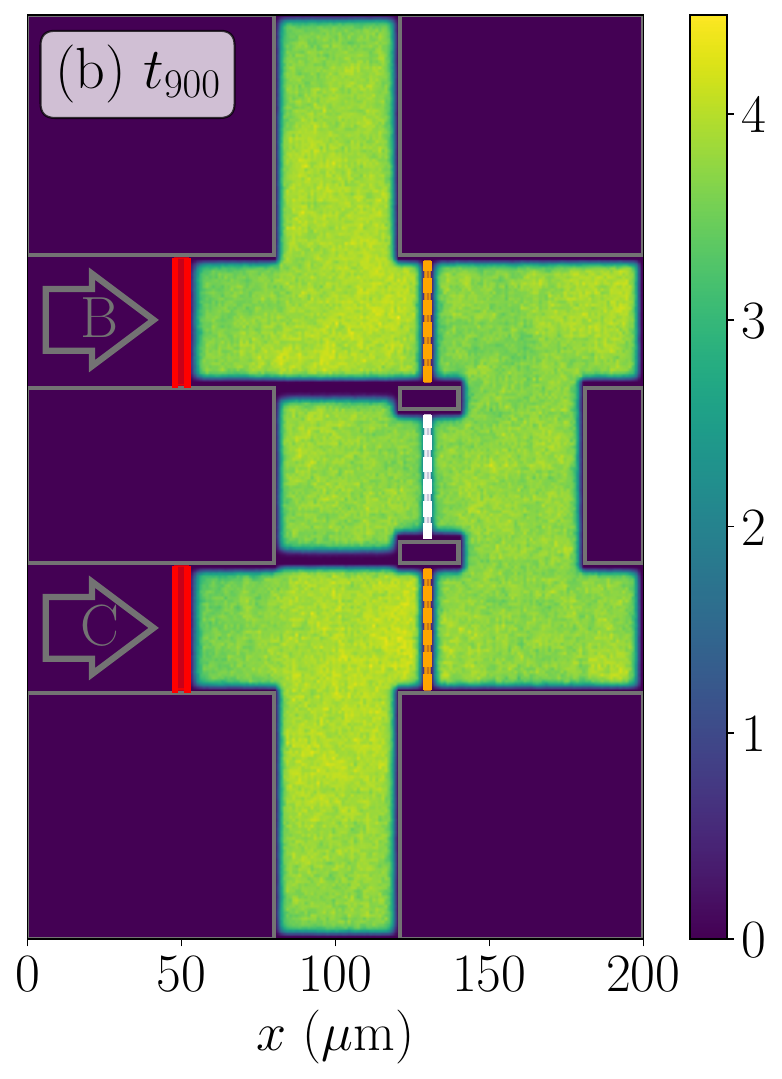}\hspace{-0em}
        \includegraphics[height=7.3cm]{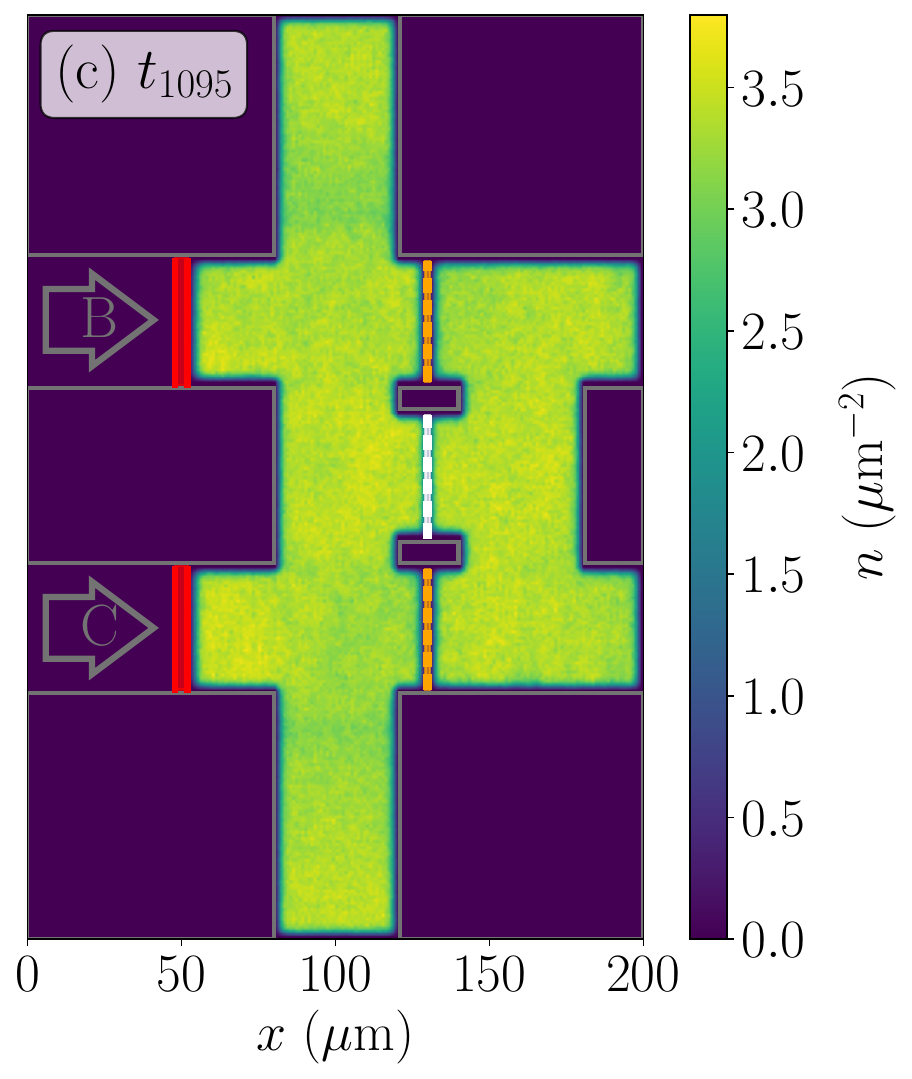}
          
    \caption{Dynamical regimes of an atomtronic circuit. 
    (a) Density distribution $n(x,y)$ is shown at time $t=650\text{ ms}$, after only the horizontal barriers B and C are moved together at a constant velocity of $v_0= 0.1 \text{ mm/s}$ from the edges of the condensate to their final positions. Each two-input AND gate consists of two mobile barriers with heights $\tV_{0, \mm}=3$ and widths $\tsigma_\mm= 8$, and one static barrier with a height $\tV_{0, \ms}=2.5$ and width $\tsigma_\ms= 3$ (orange dashed lines).  
    Four Gaussian barriers, each with a height of $\tV_{0, \ms}=5$ and a width of $\tsigma_\ms = 3$, separate the two-input AND gates from the central part,  functioning as left and right output gates. A static barrier with a height of $\tV_{0, \ms}=2$ and width $\tsigma_\ms= 2$ (white dashed lines) is placed in the central part, acting as the output channel. 
    (b)  At $t=900\text{ ms}$, $n(x,y)$ is shown after the right output gates are opened. (c) At $t=1095\text{ ms}$,  $n(x,y)$ is displayed after the left output gates are opened. }
    \label{fig3:Dynamical Regimes of an atomtronic 4-input AND-gateEXTRA}
\end{figure*}

\section{\label{sec:UniversalSetOfGate}Universal Set of Gates}

In the previous sections, we have demonstrated how a T-shaped BEC can be utilized to create an atomtronic AND gate, and how these gates can be operated sequentially to create circuits. 
Now, we extend this concept by constructing a logical NOT gate. 
To achieve this, we create an atomic JJ by separating a 2D cloud into two subsystems using a static tunnel barrier. This approach is inspired by experiments that observe Josephson oscillations, triggered by an instantaneous phase imprint on one of the subsystems \cite{Luick2020}.  
We use a rectangular cloud with dimensions $L_x\times L_y = 40\times 40\,\mu \text{m}^2$. 
The tunnel barrier has a height of $\tV_{0, \ms} =1.5$ and a width of $\tsigma_\ms =3$. 
At time $t_\text{phase}$, we imprint a phase of $\phi_0=0.25\,\pi$ on the left subsystem, introducing a phase difference between the two subsystems. 
This phase difference excites Josephson oscillations in the density imbalance between them. 
In Fig. \ref{fig:NOT/NAND-gate Symbol + Table}(b) we display $z(t)$ after the phase imprint, for half a Josephson oscillation (JO). 
During the first quarter of the JO, the imbalance grows from zero to a maximum value, representing the NOT-gate operation where the input is zero and the output is nonzero.  In the second quarter of the JO, the imbalance reverses, representing the NOT-gate operation where the input is nonzero and the output is zero. 
These results are summarized in the table in Fig. \ref{fig:NOT/NAND-gate Symbol + Table}(a).

This NOT-gate implementation is fully compatible with the 2-input AND gate. 
By combining the NOT gate with the AND gate, we create a complete set of logical gates, allowing us to realize a NAND gate. For this purpose, we use the same system parameters as in Sec. \ref{sec:2InputAndGate} to integrate the AND and NOT gates into a complementary NAND gate.
The barrier protocol proceeds as follows: We first run the 2-input AND-gate protocol similarly to Sec. \ref{sec:2InputAndGate}, with a slight modification—here, we move the barriers by $\Delta x = \Delta y = 65 \,\mu \text{m}$, and the static barrier is shifted
to $x_{0, \ms}=145 \,\mu \mm$ to accommodate the $40\times 40 \, \mu \mm^2$ enclosure for the NOT gate. 

Next, to define the NOT-gate area, we introduce two additional static barriers, each with a height of $\tV_{0, \ms}=5.0$ and a width of $\tsigma_\ms = 3$, at $x_{0, \ms}=125\,\mu \mm$ and $x_{0, \ms}=165\,\mu \mm$. 
During this period, the original static barrier is ramped up again, 
increasing from $\tV_{0, \ms} =2.5$ to $\tV_{0, \ms} =5$, restricting the output to a smaller region within the NOT-gate area. 
After the waiting period, this static barrier is ramped down to $\tV_{0, \ms} =1.5$ over $50$ ms, which is the barrier height for the NOT gate. 
At this point, we imprint a phase of $\phi_0 =0.25 \, \pi$ on the left subsystem to generate the desired output for the NAND gate. 
In Fig. \ref{fig:NAND}, we show three cases of input barriers before and after the phase imprint.  
As expected, the density imbalance observed in Fig. \ref{fig:NAND}(a)—when both barriers are moved in the AND gate—is no longer measurable in Fig. \ref{fig:NAND}(d) after applying the NOT-gate protocol. However, if only one barrier is moved [see Figs. \ref{fig:NAND}(b) and \ref{fig:NAND}(c)], there is no resulting density imbalance. After the phase imprint, a nonzero imbalance appears in Figs. \ref{fig:NAND}(e) and \ref{fig:NAND}(f).
We summarize the results of the NAND gate in Fig. \ref{fig:NOT/NAND-gate Symbol + Table}.

\section{\label{sec:Conclusions}Conclusions}

In conclusion, we have demonstrated how to create atomtronic circuits out of Bose-Einstein condensates. As a first example and building block of circuits, we have demonstrated how to use a T-shaped condensate to create an AND gate. The logical input of the gate is given by whether or not Gaussian barriers are pushed into two of the three arms of the T-shaped condensate, to create currents along them, in a piston-like fashion. To create a logical output of the gate, we place a static tunnel barrier in the third arm of condensate. The velocities of the two input barriers is chosen such that if both are moving simultaneously, the current at the tunnel junction is larger than the critical current. As a result, a density imbalance builds up. However, if only one of the barriers is moving, the induced current is below the critical current, and thus no density imbalance builds up. Therefore, the density imbalance across the static barrier serves as the logical output of the AND gate.
Going beyond a single gate, we have demonstrated how to operate AND gates in sequence, to create an atomtronic circuit.
Furthermore, we have put forth a proposal of how to create a NOT gate in a condensate.
For this, we utilize Josephson oscillations. Specifically, we propose to use a quarter of a Josephson oscillation, which transforms a maximal density imbalance into no imbalance, and no imbalance into a maximal imbalance across the tunnel junction.
With these circuit elements, AND gate, NOT gate, and how to operate these gates in sequence, we have
demonstrated the construction of a NAND  gate, thus providing a universal set of logic gates capable
of performing any classical logical operation.

\begin{acknowledgments}
This work is funded by the Deutsche Forschungsgemeinschaft (DFG, German Research Foundation) -SFB-925 - project 170620586 and the Cluster of Excellence ’Advanced Imaging of Matter’ (EXC 2056) project 390715994. The project is co-financed by ERDF of
the European Union and by ’Fonds of the Hamburg Ministry of Science, Research, Equalities and Districts (BWFGB)’.
\end{acknowledgments}

\appendix

\section{4-Input AND gate}
\label{sec:appA}
\noindent 

\begin{table}[t!]
    \centering
    \begin{tabular}{c|c|c|c|c}
      \hline
      \,\,\,\,\,\,$v_A$\,\,\,\,\,\, & \,\,\,\,\,\,$v_B$\,\,\,\,\,\, & \,\,\,\,\,\,$v_C$\,\,\,\,\,\, & \,\,\,\,\,\,$v_D$\,\,\,\,\,\, & \,\,\,\,\,\,$z(t)$\,\,\,\,\,\, \\
      \hline \hline
      $0$ & $0$ & $0$ & $0$   & $0$ \\
      $0$ & $0$ & $0$ & $v_{0}$   & $0$ \\
      $0$ & $0$ & $v_{0}$ & $0$   & $0$ \\
      $0$ & $0$ & $v_{0}$ & $v_{0}$   & $0$ \\
      $0$ & $v_{0}$ & $0$ & $0$   & $0$ \\
      $0$ & $v_{0}$ & $v_{0}$ & $0$   & $0$ \\
      $0$ & $v_{0}$ & $v_{0}$ & $v_{0}$   & $0$ \\
      $0$ & $v_{0}$ & $0$ & $v_{0}$   & $0$ \\
      $v_{0}$ & $0$ & $0$ & $0$   & $0$ \\
      $v_{0}$ & $0$ & $0$ & $v_{0}$   & $0$ \\
      $v_{0}$ & $0$ & $v_{0}$ & $0$   & $0$ \\
      $v_{0}$ & $0$ & $v_{0}$ & $v_{0}$   & $0$ \\
      $v_{0}$ & $v_{0}$ & $0$ & $0$   & $0$ \\
      $v_{0}$ & $v_{0}$ & $0$ & $ v_{0}$   & $0$ \\
      $v_{0}$ & $v_{0}$ & $v_{0}$ & $0$   & $0$ \\
      $v_{0}$ & $v_{0}$ & $v_{0}$ & $v_0$   & $> 0$ \\
      \hline
    \end{tabular}
    \caption{Logical truth tables of 4-Input AND-gate circuit. Here we depict all 16 cases corresponding to the circuit in Fig. \ref{fig4}. $v_A,\,v_B,\,v_C$ and $v_D$ denote the barrier velocities of the horizontal and vertical moving barriers which can be moved at constant velocity $v_0 = 0.1$ mm/s (compare with the dynamical regime in Fig. \ref{fig3} and Fig. \ref{fig3:Dynamical Regimes of an atomtronic 4-input AND-gateEXTRA}).}
    \label{tab:4AND}
\end{table}

\noindent

\noindent

As a corresponding counterexample to the discussed case of the 4-input AND-gate in Sec. \ref{sec:4InputAndGate}, Fig. \ref{fig3}, where all four barriers are moved, we will now return to the case where only the two horizontal barriers are moved. This is one of a total of 16 cases where no notably large density imbalance $z(t)$ is measured as an output, see Tab. \ref{tab:4AND}.

In Fig. \ref{fig3:Dynamical Regimes of an atomtronic 4-input AND-gateEXTRA}, we show the case of moving the two horizontal mobile barriers, depicted at the three most significant transition points: After the barriers have been moved to the end position (Fig. \ref{fig3}, after the right Gaussian output barriers, which represent the link to the last AND-gate, have been removed (Fig. \ref{fig3:Dynamical Regimes of an atomtronic 4-input AND-gateEXTRA}b) and after the left output gates have been opened (Fig. \ref{fig3:Dynamical Regimes of an atomtronic 4-input AND-gateEXTRA}c). In contrast to the case in Fig. \ref{fig3}, the shifting of the barriers should not generate an output that can be quantified, i.e. the density imbalance calculated around the stationary barrier according to Eq. \ref{eq:BarrierPotentialStat} should approach zero. In this scenario, it can be observed that the density distribution of the BEC can reach an equilibrium in all three steps via Josephson tunnelling processes and the geometrical setup.

\newpage
\bibliography{paper}

\end{document}